\documentclass[aps,prd,superscriptaddress]{revtex4-1}
\usepackage{graphicx}
\usepackage{amssymb}
\usepackage{amsmath}
\usepackage{bm}
\usepackage{verbatim}

\newcommand{\dert}{\partial_t}
\newcommand{\derx}{\partial_x}
\newcommand{\der}{\partial}

\begin{document}

\title{Dynamics and Bloch oscillations
of mobile impurities in one-dimensional quantum liquids }

\author{M. Schecter}
\affiliation{
School of Physics and Astronomy, University of Minnesota,
Minneapolis, MN 55455}

\author{D.M.~Gangardt}
\affiliation{School of Physics and Astronomy, University of Birmingham,
Edgbaston,
Birmingham, B15 2TT, UK  }

\author{A. Kamenev }
\affiliation{William I. Fine Theoretical Physics Institute and
School of Physics and Astronomy, University of Minnesota,
Minneapolis, MN 55455}

\date{\today}

\begin{abstract}
  We study dynamics of a mobile impurity moving in a one-dimensional quantum
  liquid.  Such an impurity induces a strong non-linear depletion of the
  liquid around it.  The dispersion relation of the combined object, called
  depleton, is a periodic function of its momentum with the period $2\pi n$,
  where $n$ is the mean density of the liquid. In the adiabatic approximation
  a constant external force acting on the impurity leads to the Bloch
  oscillations of the impurity around a fixed position. Dynamically such
  oscillations are accompanied by the radiation of energy in the form of
  phonons.  The ensuing energy loss results in the uniform drift of the
  oscillation center. We derive exact results for the radiation-induced
  mobility as well as the thermal friction force in terms of the equilibrium
  dispersion relation of the dressed impurity (depleton).  These results show
  that there is a wide range of external forces where the (drifted) Bloch
  oscillations exist and may be observed experimentally.
\end{abstract}
\maketitle

\section{Introduction}
\label{sec:intro}

Recent experiments
\cite{Koehl_PhysRevLett.103.150601,Zipkes10,
Schmid_etal_PhysRevLett.105.133202,Wicke_2010arXiv1010.4545W}
achieved a substantial progress in fabricating and studying dilute impurities
immersed in one-dimensional (1d) quantum liquids.  Such liquids are formed by
ultracold bosonic or fermionic atomic gases placed in 1d optical lattices
\cite{Moritz_PhysRevLett.91.250402,Stoeferle_etal_PhysRevLett.92.130403,Fertig_etal_PhysRevLett.94.120403,Kinoshita04,Kinoshita06}
or using the magnetic confinement on atom chips
\cite{Wicke_2010arXiv1010.4545W}.  
A hyperfine state of a few atoms is then switched locally
with the help of  the RF magnetic field pulse
\cite{Koehl_PhysRevLett.103.150601,Wicke_2010arXiv1010.4545W}. As a result, 
the  atoms become  distinguishable from the rest of the quantum liquid and
thus may be considered as mobile impurities. Another promising realization is
achieved by placing \emph{ions} of  $\mathrm{Yb}^+$, $\mathrm{Ba}^+$  or
$\mathrm{Rb}^+$   
in the Bose-Einstein condensate of neutral ${}^{87}\mathrm{Rb}$ 
atoms \cite{Zipkes10,Schmid_etal_PhysRevLett.105.133202}.

Remarkably, the impurities may be selectively acted upon with the help of
external forces.  In the case of neutral atoms the external force is
gravity,  uncompensated by the force from the vertical magnetic trap
as the impurity atoms are created in magnetically untrapped states. 
In the case of an ion, the selective force is due to the applied electric
field. These setups thus allow to drive the 
non-equilibrium dynamics of mobile impurities immersed in a 1d quantum liquid.  
By releasing the trap after a certain delay time allows 
monitoring the resulting density and velocity distribution of impurities.
\begin{figure}[b] \centering
  \includegraphics[width=0.55\columnwidth]{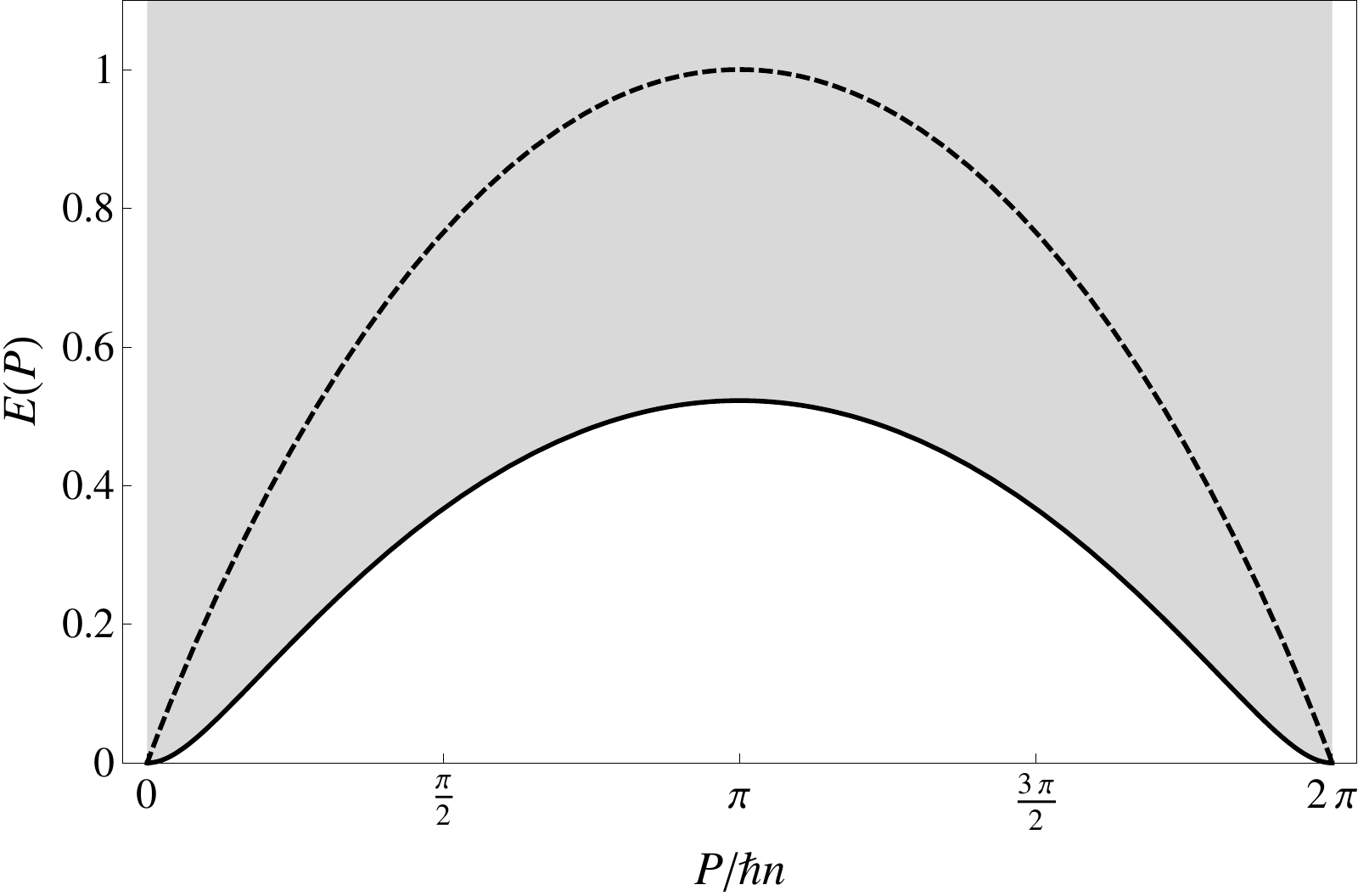}
  \caption{Schematic dispersion relation for a mobile  impurity in a weakly interacting 
    superfluid.  The dispersion defines the lower bound of the many-particle
    excitation spectrum. In the region above the dispersion (shaded gray)
    there exists a continuum of many-body states. The energy is measured  
    in units of the maximum energy of soliton excitations in the superfluid
    background, shown by the dashed line.}
\label{fig:dispersion}
\end{figure}

It is important to emphasize that in this publication we are dealing with the
case of impurities with finite bare mass $M$, not very different from the mass
$m$ of the particles in the background. The velocity of the impurity is
therefore free to change as a result of absorbed momentum.  This situation is
qualitatively different from that of static, infinitely massive impurities 
\cite{Kane_Fisher_PhysRevB.46.15233,Hakim_PhysRevE.55.2835,Buechler_Geshkenbein_Blatter_2001,Astrakharchik2004Motion}.

Dynamics of mobile impurities in quantum liquids is an old subject pioneered by
Landau and Khalatnikov
\cite{LandauKhalatnikov1949ViscosityI,LandauKhalatnikov1949ViscosityII} and
Bardeen, Baym, Pines and Ebner
\cite{Bardeen_etal_PhysRevLett.17.372,*Bardeen_etal_PhysRev.156.207,*Baym_PhysRevLett.17.952,*Baym_PhysRevLett.18.71,*Ebner_PhysRev.156.222,BaymEbner1967Phonon}
in their studies of ${}^3\mathrm{He}$ atoms in superfluid ${}^4\mathrm{He}$.
These authors realized that at a finite temperature the impurities experience
a viscous force from the normal component of the liquid, even if their speed
is less than the critical superfluid velocity. The mechanism leading to such a
viscous force was shown to be the Raman scattering of the thermal excitations
of the superfluid \emph{i.e.}  thermal phonons scattering off the impurity
${}^3\mathrm{He}$ atoms.  It was found
\cite{LandauKhalatnikov1949ViscosityI,*LandauKhalatnikov1949ViscosityII,BaymEbner1967Phonon}
that the corresponding friction coefficient scales with temperature as $T^8$
in the low temperature limit.  It was later discussed by Castro-Neto and
Fisher \cite{castro96} that in 1d the Raman mechanism leads to the friction
force proportional to $T^4$ due to the phase-space considerations. The same
mechanism governs the temperature-induced acceleration of grey solitons
\cite{Muryshev2002Dynamics}.  Recently,
two of the present authors \cite{Gangardt09,Gangardt2010Quantum} 
showed that in addition to
phase-space effects the friction is extremely sensitive to the details of the
interactions between impurity and the background liquid and vanishes for
exactly solvable 1d models.  The friction therefore may be rather small in the
spin-flip setup if the parameters are near the Yang-Gaudin
\cite{CN_Yang_1967,*Gaudin_1967} integrable case.

Due to the one-dimensional kinematics
the slowly moving impurity tends to deplete the host liquid  creating
density and current gradients in its vicinity. At
small momenta the dispersion of the impurity  can still be described by 
quadratic law $E=P^2/2M^*$ with effective mass $M^*$ of the impurity atom
{\cite{Fuchs_2005,Matveev08}}. In the weakly interacting  
regime and sufficiently small momenta the depletion of the host liquid
can be studied within the framework of  Bogoliubov theory. 
The resulting excitations, polarons,
consist of  impurities surrounded by a cloud of linear 
excitations of the condensate  and were studied in
Ref.~\cite{Lee_Gunn_PhysRevB.46.301} and later in Refs.~\cite{Cucchietti_Timmermans_PhysRevLett.96.210401,Bruderer_etal_PhysRevA.76.011605}.  
However, even for a weakly interacting liquid, the induced depletion becomes
essentially \emph{non-linear} at larger momenta and/or strong impurity-liquid
coupling. This brings a qualitative change of the dispersion relation and
we introduce a quasiparticle which we call {\em depleton}
describing the impurity dressed by the non-linear 
depletion cloud.

The remarkable
property of depleton dispersion is that it is a {\em periodic} function of the
momentum $P$ with the period $2\pi\hbar n$, 
where $n$ is the density of 1d host liquid.
To explain this feature it is worth noticing that, being quadratic at small
momenta, the depleton energy  is less than any of the
excitations of the host liquid (see Fig.~\ref{fig:dispersion}). 
Indeed, the low-energy excitations of the host liquid have a sound-like nature
with the energy $cP>P^2/2M^*$, where $c$ is the speed of sound. Therefore the
dressed impurity excitation provides the cheapest way for the system to
accommodate a small momentum.  This means that the depleton dispersion
relation  is {\em defined} as the lowest possible many-body excitation
energy of the system with a given momentum. An example of such dispersion is
shown in Fig.~\ref{fig:dispersion}. Above the depleton  
energy there is a continuum of  many-body excitations 
comprised of the moving impurity and a
certain number of phonons. This spectral edge is characterized by the
power-low singularities of zero-temperature correlation functions 
and was  discussed in
Refs.~\cite{zvonarev_2007,Kamenev08,Matveev08,Zvonarev_etal_PhysRevB.80.201102}.

One may argue now that in the infinite system the ground-state energy with a
given momentum is a periodic function of the latter with the period $2\pi\hbar
n$. Indeed, it is easy to see that the ground-state energy with momentum 
$2\pi\hbar n$ vanishes in the thermodynamic limit.  
To this end consider a ring of length $L$  where the spectrum of the momentum
operator is quantized in units of $2\pi\hbar/L$. 
If the momentum of each particle is boosted by one quantized unit, 
the total momentum of the system is $2\pi\hbar  n$, 
while the total energy vanishes thanks to the total mass diverging in the
thermodynamic limit.  We thus conclude that the ground-state for a given
momentum $P$ which is the dispersion relation $E(P)$ of the dressed impurity
is a periodic function of momentum. Explicit examples are provided by exactly
solvable models \cite{zvonarev_2007} and weakly interacting Bose-liquid 
corresponding to Fig.~\ref{fig:dispersion} and 
considered in details in Appendix~\ref{sec:equilibrium impurity}.

The physics behind the periodic dispersion relation is in the transfer of
momentum from the accelerated impurity to the supercurrent in the background
liquid: similarly  to  the density depletion,
the moving impurity creates the sharp phase drop $\Phi$ across it. To satisfy
the periodic boundary conditions the rest of the liquid must sustain the
phase gradient $\Phi/L$, resulting in the supercurrent 
which carries momentum $\hbar n \Phi$. 
While the supercurrent is absorbing
momentum, it does not contribute to energy. 
Indeed, as it was already mentioned, in the thermodynamic limit the
bulk of the liquid is infinitely heavy and thus can accommodate any momentum
at no energy cost.  As a result, the energy and momentum of the
depleton core, being periodic functions of the phase drop $\Phi$ with the
period $2\pi$, oscillate as functions of the total momentum with the period
$2\pi \hbar  n$, while the rest of the momentum goes into the supercurrent.
A similar periodic dispersion relation would arise if the host
liquid were considered as a rigid crystal with the lattice constant
$a=1/n$. Then the momentum interval between $P=-\hbar\pi n$ and 
$P=\hbar\pi n$ is nothing but the Brillouin zone of such a crystal, 
while the impurity dispersion, discussed above, is its lowest Bloch band.

Although the above considerations completely disregard the absence of the true
long-range order in the 1d liquid (either superfluid or crystalline) and thus
should be taken with care, the periodicity of the dispersion suggests the
possibility of Bloch oscillations of the impurity atom subject to an external
force \cite{Gangardt09}.  Indeed, if an external force $F$ is applied to the
impurity atom (\emph{e.g.} electric field is acting on an ion) the {\em total}
momentum of the system changes linearly with time, $P=Ft$. For infinitesimal
force this leads to adiabatic change of the energy $E(P)$ and the velocity
$V=\partial E/\partial P$ of the depleton which become periodic functions of
time with the period $2\pi\hbar n/F$. It is quite remarkable that in such a
process, on average, the impurity does not accelerate; moreover, it does not
even move.  Instead, it channels the momentum into the collective motion,
\emph{i.e.}, the supercurrent, of the liquid and in the process oscillates
around a fixed location.  As a result, {\em no energy} is transferred, on
average, to the system from the external potential.

This spectacular phenomenon is present at zero temperature and under an
infinitesimal external force. Both finite temperature and a finite force
complicate the picture in a substantial way. The aim of this paper is to
clarify the influence of these two factors on the observability of the Bloch
oscillations. In brief, our conclusions are as follows: at a sufficiently
low temperature there is a parametrically wide range of external forces
$F_\mathrm{min}(T)<F<F_\mathrm{max}$, where the Bloch oscillations {\em are
  observable}. Contrary to the adiabatic picture, they are accompanied by a
{\em drift} and exhibit certain amplitude and period renormalization.

The drift manifests itself in the appearance of an average velocity $V_D$,
superimposed on top of the periodic  Bloch oscillations. It is  a 
linear function of force at small
forces, allowing to define the impurity {\em mobility} $\sigma$, as $V_D=
\sigma F$.  As a result, the total energy of the system increases (\emph{i.e.}
the system is heated) with the rate $FV_D = \sigma F^2$.  This energy goes to
the emitted long wavelength phonons, which run away from the impurity with the
sound velocity $c$. The maximal force can then be estimated as $F_\mathrm{\max}
\approx c/\sigma$.  
At a larger force the drift velocity exceeds $c$, leading
to Cherenkov radiation of phonons. The phonons take a substantial
part of the momentum and thus ruin the Bloch oscillation mechanism, discussed
above. The impurity motion is then either incoherent drift, or an unlimited
acceleration, depending on the parameters.

We derive the {\em exact} analytic expression for the drift
mobility $\sigma$ expressed in terms of the equilibrium dispersion relation of
the impurity. It is worth noticing that the mobility $\sigma$ is {\em
  not} the linear response property, despite the linear relation
between the drift velocity and the external force.  Indeed, the dynamics of
depleton in this regime is the drift
superimposed with the essentially non-linear pattern
of Bloch oscillations in which  the particle  explores the entire 
range of impurity momenta and energy. 
It is therefore not immediately obvious that the mobility
$\sigma$ may be expressed in terms of the equilibrium properties. Nevertheless
we prove that such a relation does exist.

The true linear response is associated with the thermally induced
Landau-Khalatnikov friction force $F_\mathrm{fr} (V)\propto -T^4 V$, which
arises due to the Raman scattering of phonons discussed above.  It provides
the lower limitation on the externally applied force $F_\mathrm{min}(T)=
F_\mathrm{fr} (V_c)$, where $V_c$ is the maximal equilibrium velocity given by
the maximum slope of the depleton dispersion,
Fig.~\ref{fig:dispersion}. Indeed, at smaller external forces the velocity
saturates and therefore the depleton dynamics is confined to the small momenta
and Bloch oscillations do not occur.

The friction coefficient in the small momentum regime
was discussed in Ref.~\cite{Gangardt09} and found to be vanishing in exactly
solvable cases. The reason behind it is the presence of infinite number of the
conservation laws, which prevent a non-equilibrium state from thermalization.
Here we extend those calculations to the case where the
impurity explores the entire range of momenta and derive an exact result for the
full momentum dependence of the friction force. Quite naturally, it vanishes
in exactly solvable cases too.

The outline of the paper is as follows: in 
Section~\ref{sec:qualiative-analysis} we give a qualitative description of the
depleton quasiparticles and the mechanism of  coupling to linear sound 
excitations, or phonons,\ of the background liquid.  
Section \ref{sec:QL} is devoted to the formal derivation of the depleton
Lagrangian based on superfluid thermodynamics. 
We discuss coupling of the depleton with
the phonon subsystem and derive a set of stochastic equations of motion for the
impurity dynamics in Section \ref{sec:phonon_coupling}.  These equation are then
used to  study radiation losses and derive expression of depleton 
mobility in Section \ref{sec:coupling_phonons} and thermal friction in Section
\ref{sec:noise}. The main results are summarized in Section
\ref{sec:discussion}. Technical details are delegated to 
Appendices.

\section{Qualitative analysis}
\label{sec:qualiative-analysis}

The key to understanding the impurity dynamics is in its interactions with the
phonons of the host liquid.  This problem is rather non-trivial even if the
impurity is weakly coupled to the liquid. Indeed, no matter how weak the
interactions are, the impurity develops local depletion, which become
appreciable when the impurity momentum approaches $\pi n$ (we set $\hbar=1$
throughout the rest of the  paper). To visualize this process it is useful to 
assume a semiclassical picture of the background, valid for weakly
interacting Bose liquid. In this regime, to accommodate the total momentum
$\pi n$ the
depletion cloud takes the form of the dark soliton \cite{Tsuzuki_1971}.  The
dark soliton is an essentially non-linear mesoscopic object, which includes a
large number of particles and a complete depletion of the liquid density.  It
is exactly the soliton formation which is responsible for channeling momentum
into the supercurrent and thus for the Bloch oscillations. It is also the
soliton which determines the interactions of the impurity with the dynamically
induced phonons. Therefore the non-equilibrium dynamics of the quantum
impurity cannot be separated from the dynamics of the essentially non-linear
soliton-like depletion cloud.

What makes the problem analytically tractable is the scale separation between
the spatial extent of the local soliton-like cloud and the characteristic
phonon wavelength. The former is given by the healing length $\xi= 1/mc$. 
The latter appears to be much longer than $\xi$, if the temperature is
sufficiently low  and the external force is not too
strong. One can thus separate the near-field
mesoscopic region, which contains the quantum impurity and its depletion
cloud, from the far-field region, supporting the radiation emitted by the
impurity.  Since the depletion is restored  exponentially at the healing
length away from the impurity, the precise position of the boundary between
the near-field and the far-field regions is of no importance.

Because of the  wide difference in their spatial scales, 
the impurity together
with its entire non-linear depletion cloud represents a dynamic {\em
  point-like} scatterer for the long wavelength phonons. From the viewpoint of
such phonons any point scatterer may be entirely described by two phase
shifts. These two phase shifts are the discontinuities of the phonon
displacement and momentum fields across the scatterer. They may be expressed
through the number of depleted particles $N$ and the phase drop $\Phi$ across
the depleton quasiparticle.  Therefore out of many degrees of freedom of the
near-field region only $N$ and $\Phi$ 
interact with the phononic sub-system.

What remains is to describe the dynamics of the local depletion cloud with
certain fixed values of $N$ and $\Phi$.  Solution of this latter problem is
facilitated by the fact that the characteristic equilibration rate of the
cloud, estimated as  $1/\tau=c/\xi$, 
is much faster than the relevant phononic frequencies.  As a
result, the cloud may be treated as being in the state of {\em local
  equilibrium}, conditioned to certain values of the slow collective
coordinates $N$ and $\Phi$. The fact that there are two such slow variables is
due to the presence of the two conservation laws: number of particles and
momentum. The fast internal equilibration of the near-field region is
therefore conditioned by the instantaneous values of the two conserved
quantities.

The slow change (compared to the fast time scale of $\tau$) in the number of 
depleted particles $N$ and the phase drop
$\Phi$ are due to the fact that the {\em local} chemical potential $\mu$  and
the local current $j$ at the position of the impurity are both affected by the
state of the global phononic sub-system. As a result, one may express the
Lagrangian of the near-field region as a function of $N$ and $\Phi$ through
the equilibrium thermodynamic potential which is a function of 
$\mu$ and $j$. 
This latter function may be independently measured or analytically
evaluated in certain limiting cases and for exactly solvable models.

These considerations allow one to separate the local, non-linear but {\em
  equilibrium} problem, from the global, non-equilibrium but {\em linear} one.
The latter statement implies that the host liquid sufficiently far away from
the impurity may be treated as the linear \emph{i.e.} 
Luttinger liquid \cite{PopovBookFunctional,HaldanePRL81}. 
This is certainly an approximation which
disregards the possibility of the moving impurity to emit non-linear
excitations, such as grey solitons or shock waves.  
The train of solitons emitted by the
impurity moving with a constant supercritical velocity was indeed observed in
simulations of Ref.~\cite{Hakim_PhysRevE.55.2835}.  The kinematics of this
process suggests that it is only possible if the drift velocity is close to
the speed of sound $c$. We therefore assume that as long as $F<
F_\mathrm{max}$, one may disregard solitons emission and treat the
liquid away from the impurity as the linear one. This is essentially the same
criterion, which allows us to separate the depletion cloud from the long
wavelength phonons.

Adopting these approximations, one is able to integrate out the phononic
degrees of freedom, characterizing the liquid away from the impurity.  It
reduces the problem to the dynamics of the impurity described by its
coordinate $X(t)$ and momentum $P(t)$ along with the dynamics of its
near-field depletion cloud fully described by the two collective coordinates
\emph{i.e.} number of depleted particles $N(t)$ and the phase drop
$\Phi(t)$. We derive an effective  action written in terms of
such an extended set of degrees of freedom. Such an action leads to the
coupled system of quantum Langevin equations governing the dynamics of the
depleton.

Away from equilibrium, for $F> F_\mathrm{min}$, the equations of motion 
yield the
pattern of Bloch oscillations.  
The deterministic part of these equations
provides with the information about drift velocity,
amplitude and shape of the velocity oscillations, as well as their period.
The stochastic part results in a certain dephasing of the oscillations. 
It is interesting to notice that the stochastic part is manifestly 
different from the equilibrium noise,
prescribed by the fluctuation-dissipation theorem. As a
consequence, the exactly integrable models loose their special status and
their non-equilibrium dynamics appears to be not qualitatively 
different from the
dynamics of generic non-integrable models.

\section{Lagrangian of the mobile impurity}
\label{sec:QL}

Let us first consider the background liquid in the absence of impurity
employing hydrodynamical description proposed by Popov
\cite{PopovBookFunctional}. Its Lagrangian is expressed in terms of the slowly
varying chemical potential $\mu$ and density $n$ as an integral of the local
thermodynamic pressure
\begin{eqnarray}
  \label{eq:grand-canonical_prime} L_0 (\mu,n) = \int\! dx\, p_0(\mu, n) =
\int\! dx\,\Big[\mu n-e_0(n)\Big].
\end{eqnarray}
Here $e_0(n)$ is the energy density of the liquid.
In the thermodynamic equilibrium the density is a function of
the chemical potential, given by a solution of the following equation:
$\mu =\mu(n)=
\der e_0/\der n$, which is a result of the minimization of this functional with
respect to $n$. This way one defines the grandcanonical thermodynamic
potential of the host liquid as
\begin{eqnarray}
  \label{eq:grand-canonical} \Omega_0(\mu) = - L_0 (\mu,n(\mu))\, .
\end{eqnarray}
For a uniform system the Lagrangian and the corresponding
thermodynamic potential are both proportional to the length of the system.

\begin{figure}[t] \centering
  \includegraphics[width=0.7\columnwidth]{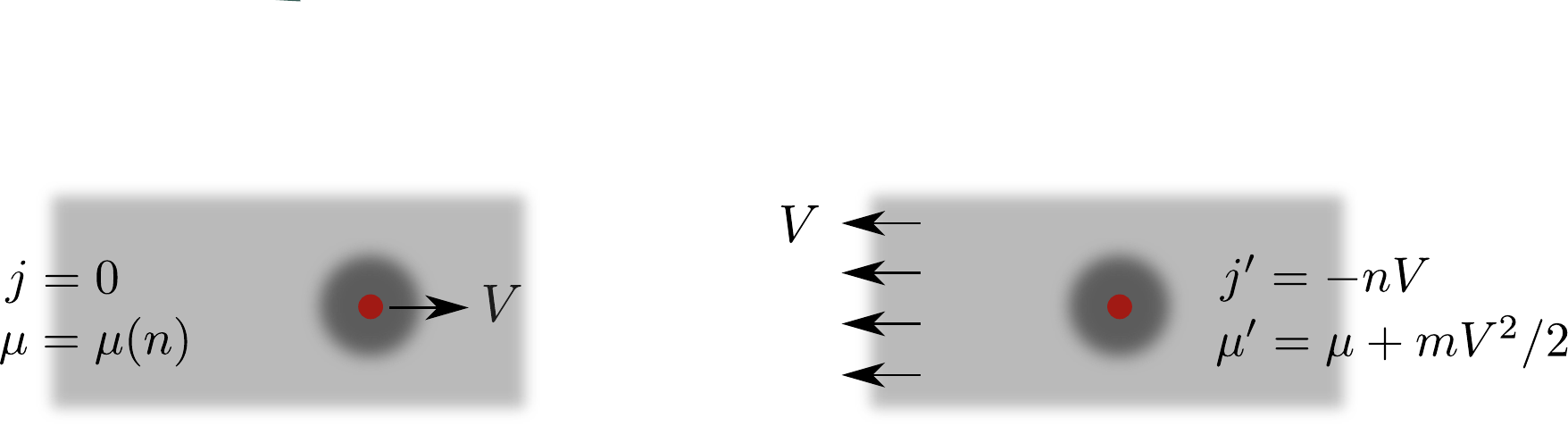}
  \caption{Transformation to moving frame.}
  \label{fig:frame}
\end{figure}

Consider now an impurity of mass $M$, having a coordinate $X$ and moving
through the liquid with velocity $V=\dot X$, as measured in the laboratory
reference frame.  It is convenient to choose the reference
frame where the impurity is at rest and the liquid flows with the velocity
$-V$, as shown in Fig.~\ref{fig:frame}.  In this co-moving frame the impurity
experiences the supercurrent $j'$ and the chemical potential $\mu'$.
Hereafter primes denote physical quantities defined in the co-moving
frame to distinguish them from the corresponding quantities in the laboratory
frame. We employ Galilean transformation into the moving frame which gives
\begin{eqnarray}
  \label{eq:jmu}
  j'=-nV,\qquad
  \mu' = \mu +mV^2/2 .
\end{eqnarray}
Together with the Galilean transformation of the energy density
$e'_0 = e_0+ mV^2/2$, the transformation (\ref{eq:jmu}) combined with
Eqs.~(\ref{eq:grand-canonical_prime}) and (\ref{eq:grand-canonical})  show
the invariance of the background grandcanonical potential
$\Omega'_0(\mu') = \Omega_0(\mu)$. As expected from the Galilean invariance, the
latter is independent  of the velocity $V$ or 
the supercurrent $j'$. 

We introduce now the impurity into the flowing liquid maintaining 
\emph{both $j'$ and $\mu'$ fixed} and let it equilibrate.  
Its motion distorts the host
liquid density and velocity fields,
forming the depletion cloud moving along with the impurity.
The grandcanonical potential
increases by an amount $\Omega'_\mathrm{d} (j',\mu')=E'_\mathrm{d} -\mu'
N_\mathrm{d}$, where $E'_\mathrm{d}$ and $N_\mathrm{d}$ are the corresponding
changes in energy and number of particles.  Using the Galilean invariance one
can relate the energy $E'_\mathrm{d} = E_\mathrm{d} -P_\mathrm{d} V +m
N_\mathrm{d}V^2/2$ to the energy $E_\mathrm{d}$ and momentum $P_\mathrm{d}$
induced by the moving impurity in the laboratory frame. This fact and
relations (\ref{eq:jmu}) allows one to identify, in the spirit of Popov's
approach,  the Lagrangian of
the depletion cloud with the negative change of the grandcanonical
potential
\begin{eqnarray}
  \label{eq:lagrangianjm}
L_\mathrm{d}(V,n) &=&
P_\mathrm{d} V - E_\mathrm{d}+\mu N_\mathrm{d}  = - E'_\mathrm{d} + \mu' N_\mathrm{d}  = -\Omega'_\mathrm{d}
(j',\mu') \, .
\end{eqnarray}
This relation is quite remarkable as the left hand side describes dynamics of
the polarization cloud moving with the velocity $V$, while its right hand side
is the thermodynamic quantity. The link between them comes from the Galilean
transformation, Eqs.~(\ref{eq:jmu}).   The
relation between the Lagrangian and the grandcanonical potential,
Eq.~(\ref{eq:lagrangianjm}), can be viewed as a generalization of the Popov
relation, Eq.~(\ref{eq:grand-canonical}), to the case of mobile impurities.

Two remarks are in order. First, in assuming the
thermodynamic equilibrium at nonzero supercurrent $j'$ flowing
through the impurity we rely on the superfluidity.
Second, we note that the increase in energy $E'_\mathrm{d}$, momentum
$P_\mathrm{d}$,  number of particles $N_\mathrm{d}$ and the
grandcanonical potential $\Omega'_\mathrm{d}$ due to the presence of 
\emph{one single impurity} are finite size corrections to the corresponding extensive
quantities.

\subsection{Collective degrees of freedom of the depleton}
\label{sec:deg_free}

Variations of the so far fixed parameters $j'$ and $\mu'$ of the background
liquid induce changes in the thermodynamic potential of the depletion
cloud.  It can be written with the help of the corresponding response
functions as
\begin{eqnarray}
  \label{eq:thermo_der} \mathrm{d}\Omega'_\mathrm{d} = \Phi\, \mathrm{d}j' + N
\mathrm{d}\mu'\, ,\qquad \qquad\Phi=\der_{j'} \Omega'\, ,\;\;\;\;
 N=\der_{\mu'} \Omega'.
\end{eqnarray}
The response to the variation of the chemical potential
$N=-N_\mathrm{d}$ is identified with the number of particles
\emph{expelled} from the
liquid by the impurity (hence the minus sign).  The response $\Phi$ to the
change of the supercurrent $j'$ is the superfluid phase and has no analogy in
classical thermodynamics.  In the state of the global thermodynamic 
equilibrium both $\Phi$ and
$N$ are rigidly locked to $j'$ and $\mu'$ and, consequently, to
$V$ and $n$. This is denoted by writing $\Phi=\Phi_0(V,n)$ and $N=N_0(V,n)$.
These functions can be obtained from the derivatives of the Lagrangian defined
in Eq.~(\ref{eq:lagrangianjm}) as described in the next subsection.

In the nonequilibrium situations, where the supercurrent and chemical
potential fluctuate 
it is convenient to treat $\Phi$ and $N$ as
\emph{independent variables}. We perform
the standard Legendre transformation to a new thermodynamic potential,
\begin{eqnarray}
\label{eq:xi} H_\mathrm{d} (\Phi, N) = \Omega'_\mathrm{d} -j'\Phi - \mu'
N,\qquad \mathrm{d} H_\mathrm{d} = -j'\mathrm{d} \Phi - \mu'\mathrm{d}N.
\end{eqnarray}
The independent variables $\Phi$ and $N$ describe the state of the depleted
liquid in the immediate vicinity of the impurity which may or may not be in
equilibrium with globally imposed $j'$ and $\mu'$. 
In equilibrium situation $H_\mathrm{d}(\Phi,N)$ does not contain 
any additional information with
respect to the thermodynamic potential $\Omega'_\mathrm{d} (j',\mu')$, which
is in turn related to the depletion cloud Lagrangian
$L_\mathrm{d}(V,n)$ by Eq.~(\ref{eq:lagrangianjm}).
The aim of introducing $H_\mathrm{d}(\Phi,N)$ is to allow
for interactions of the depletion cloud with the long wavelength
phonons. As we shall see in Section~\ref{sec:phonon_coupling} 
the latter may change the number of particles and the momentum of the
depletion cloud, forcing it to equilibrate to some new values of $\Phi$ and
$N$.  Before turning to  phonons, it is instructive to rewrite down the
impurity Lagrangian (\ref{eq:lagrangianjm}) by substituting into it
Eq.~(\ref{eq:xi}) and considering $\Phi$ and $N$ as independent vaiables,
\begin{eqnarray}
  \label{eq:lagrangian_subst} 
  L_\mathrm{d}= \frac{1}{2}\, MV^2  -j'\Phi - \mu' N
  -H_\mathrm{d} (\Phi,N)\,.
\end{eqnarray} 
Expressing $j'$, $\mu'$ through $V$, $n$ using Eq.~(\ref{eq:jmu}), we finally
obtain the Lagrangian of the depleton
\begin{eqnarray}
  \label{eq:lagrangian_nophonons} 
  L(V,\Phi, N) = \frac{1}{2}\,
  (M-mN)V^2+nV\Phi -\mu N - H_\mathrm{d} (\Phi,N)\, ,
\end{eqnarray} 
where $\mu=\mu(n)$ is the equilibrium chemical potential of the host liquid in
the laboratory frame and we have added the bare kinetic energy proportional to
the impurity mass $M$.  The momentum of the depleton is obtained by the
standard procedure
\begin{eqnarray}
  \label{eq:momentum} P = \frac{\der L}{\der V} = (M-mN)V + n\Phi\, ,
\end{eqnarray}
The last term describes the supercurrent momentum $n\Phi$ stored in the
background.  The first term is proportional to the reduced mass $M-mN$,
expressing the fact that the mass $mN$ is {\em removed} from the local
vicinity of the moving impurity. This quantity should not be confused with the
\emph{effective} mass $M^*$ given by the curvature the equilibrium dispersion
Eq.~(\ref{eq:dispersion}) or obtained from the equilibrium dynamics in
Eq.~(\ref{eq:eff_mass}).

We can use Eq.~(\ref{eq:momentum}) to express velocity  of the
depleton as a function of its momentum,
\begin{eqnarray}
  \label{eq:velocity_phi_n}
  \dot X =V= V(P,\Phi,N) =  \frac{P-n\Phi}{M-mN}.
\end{eqnarray}
Combining Eq.~(\ref{eq:velocity_phi_n}) and the Lagrangian
(\ref{eq:lagrangian_nophonons}) leads to the Hamiltonian
\begin{eqnarray}
  \label{eq:hamiltonian_nophonons}
H(P, \Phi, N) &=& PV - L =
\frac{1}{2}\frac{(P-n\Phi)^2}{M-mN} +\mu N + H_\mathrm{d} (\Phi,N)\, .
\end{eqnarray}
This Hamiltonian generates the following equations of motion
\begin{eqnarray}
  \label{eq:p_0}
  \dot P&=&0 \\
  \label{eq:phi_0}
  0&=&- \der_\Phi H =n V -\der_\Phi H_\mathrm{d} \\
  \label{eq:n_0}
  0&=&-\der_N H =-\mu -mV^2/2  -\der_N H_\mathrm{d}
\end{eqnarray}
in addition to  Eq.~(\ref{eq:velocity_phi_n}). The first equation
is the momentum conservation expected in the absence of external forces
and for homogeneous background.
Two other equations  are in fact
\emph{static constraints}.   This is a manifestation
of the already mentioned fact that  without phonons
$\Phi$ and $N$  are rigidly locked constants
and do not have an independent dynamics.

\subsection{Dynamics of depleton in the absence of phonons and 
equilibrium  values of collective variables }
\label{sec:dyn_eq}
As we shall see in Section~\ref{sec:phonon_coupling},
the coordinates canonically conjugated to $\Phi$, $N$
are phononic displacement and phase at the location of the impurity.
In the absence of these degrees of freedom  the only consistent solution of
Eqs.~(\ref{eq:phi_0}),~(\ref{eq:n_0}) is \emph{static} relations
$\Phi=\Phi_0(P,n)$, $N=N_0(P,n)$. Substituting them back into the Hamiltonian
(\ref{eq:hamiltonian_nophonons}) leads
to the equilibrium dispersion relation of the dressed impurity (depleton),
\begin{eqnarray}
  \label{eq:dispersion}
  H(P,\Phi_0 (P,n),N_0(P,n)) = E(P,n)\, .
\end{eqnarray}
The corresponding ``equilibrium Lagrangian'' can be obtained 
as $L(V,n) = PV-E$ by expressing the momentum with the help of  
$V=\der E/\der P$. In most situations it is rather
these quantities and not the ``internal energy'' $H_\mathrm{d}(\Phi,N)$ 
represent the physical input about the dynamics of the
dressed impurity. They can be  obtained from  solving the equilibrium problem 
for the impurity moving with the constant
momentum $P$ or velocity $V$, through the liquid with the asymptotic
density $n$. Below we show explicitly how collective variables $\Phi$ and $N$
and the corresponding energy $H_\mathrm{d}$ can be obtained from the knowledge
of $E(P,n)$ or $L(V,n)$.

We start with the situation when velocity $V$ is a control parameter.
In this case finding $H_\mathrm{d}(\Phi,N)$ amounts to perform the Legendre
transformation Eq. (\ref{eq:xi}) by exploiting the definition
(\ref{eq:lagrangianjm}) of the Lagrangian $L_\mathrm{d}=L-MV^2/2$
of the depletion cloud  in terms of the thermodynamic
potential $\Omega'_\mathrm{d}$ and  using the pair $(V,n)$ instead
of the thermodynamical variables $(j',\mu')$.  To achieve this goal we
use Eqs.~(\ref{eq:jmu}) to relate the corresponding partial derivatives by the
linear transform
\begin{eqnarray}
  \label{eq:linear_derivatives}
  \begin{pmatrix} \der_V \\ \der_n\end{pmatrix} &=&
  \begin{pmatrix} - n  & mV \\ -V & mc^2/n  \end{pmatrix}
  \begin{pmatrix} \der_{j'} \\ \der_{\mu'} \end{pmatrix} .
\end{eqnarray}
Here we have used the  relation $\der \mu/\der n = mc^2/n$
between the compressibility and the sound velocity $c$.
Using the definitions Eqs.~(\ref{eq:thermo_der}) 
and Eqs.~(\ref{eq:linear_derivatives})
we can express the derivatives of $L(V,n)$ in terms of
collective variables $\Phi,N$
\begin{eqnarray}
\label{eq:dLdV}
  \frac{\der L}{\der V}-MV &=&
P-MV =
 n\Phi -  mVN \\
\label{eq:dLdn}
  \frac{\der L}{\der n} &=& 
  V \Phi - \frac{mc^2}{n}{N}.
\end{eqnarray}
Solving these equations  yield equilibrium values,
$\Phi_0(V,n)$ and $N_0(V,n)$. Eq.~(\ref{eq:dLdV}) can be otherwise
obtained by simply substituting $\Phi_0(V,n)$ and $N_0(V,n)$ into
the  definition of the momentum,  Eq.~(\ref{eq:momentum}). This is a
consequence of the equations of motion (\ref{eq:phi_0}),~(\ref{eq:n_0}). 
The quantities involving second derivatives of the Lagrangian do not enjoy
this property. The most obvious case is the effective
mass 
\begin{eqnarray}
  \label{eq:eff_mass}
M^* = \frac{\der P}{\der V}=  M-mN -m V \frac{\der N_0}{\der V} +n
\frac{\der\Phi_0}{\der V}\, ,
\end{eqnarray}
which differs from the expression $M-mN$ obtained by taking partial derivative
of Eq.~(\ref{eq:momentum}) with respect to velocity.

The equilibrium relations $\Phi_0(V,n)$ and $N_0(V,n)$ can be inverted
to find velocity  $V_0(\Phi,N)$, and density  $n_0(\Phi,N)$
for given values of $\Phi$ and $N$. Substituting them into
Eq.~(\ref{eq:lagrangian_nophonons}) gives
\begin{eqnarray}
H_\mathrm{d} (\Phi,N) = - L(V_0,n_0) +\frac{MV_0^2}{2}+
n_0V_0\Phi - \left(\mu(n_0) + \frac{mV_0^2}{2}\right)N \,.
\label{eq:lagrangian_impurity}
\end{eqnarray}

Conversely, we can use the momentum $P$ as a control parameter. 
Again, using  equations of motion  Eqs~(\ref{eq:phi_0}),~(\ref{eq:n_0})
one is able to show the equivalence of the derivatives 
$\der H(P,\Phi,N)/\der n =\der E(P,n)/\der n$ and
$\der H(P,\Phi,N)/\der P =\der E(P,n)/\der P$.  The latter defines the 
velocity $V_0(P,n)$.
Using these facts and differentiating explicitly
Eq.~(\ref{eq:hamiltonian_nophonons}) we have the following system of equations
\begin{equation}
\label{eq:Phi N}
V_0(P,n)=\frac{P-n\Phi}{M-mN};\,\,\,\,\,\quad\quad
\frac{\partial E(P,n)}{\partial n}=\frac{mc^{2}}{n}N-V_0(P,n)\Phi,
\end{equation}
which are equivalent to Eqs.~(\ref{eq:dLdV}),~(\ref{eq:dLdn}) by virtue of the
fact that $(\der L/\der n)_V = -(\der E/\der n)_P$. Solving 
Eqs.~(\ref{eq:Phi N})  
yield equilibrium values
$\Phi_0(P,n)$, $N_0(P,n)$ as functions of $P$  and $n$.  Next, we invert 
these relations to obtain
$P_0(\Phi,N), n_0(\Phi,N)$ as functions of $\Phi, N$.
Using Eq.~(\ref{eq:dispersion}) and  Eq.~(\ref{eq:hamiltonian_nophonons})
we obtain  the expression for the core energy
\begin{equation}
\label{eq:hamiltonian_impurity}
H_\mathrm{d}(\Phi,N)=
E(P_0,n_0)-\mu(n_0)N-\frac{1}{2}\frac{(P_0-n_0\Phi)^2}{M-mN}.
\end{equation}
in terms of the dispersion $E(P,n)$ and its derivatives.

To illustrate this procedure we use two cases, where the energy
$H_\mathrm{d}(\Phi, N)$ possesses a simple analytic form.  One is the grey
soliton in weakly interacting Bose-Einstein condensate describing a
massless, $M=0$, impurity propagating in a weakly interacting Bose
liquid with the coupling constant $g$. The standard results
\cite{PitaevskiiStringariBook,Tsuzuki_1971}
for the soliton
dynamics are provided in Appendix~\ref{sec:dark_soliton} and
lead to
\begin{eqnarray}
  \label{eq:h_dark_soliton} H_\mathrm{d}(\Phi,N) = \frac{1}{8}\,mg^2 N^3\left[
\frac{1}{3} - \left(\sin\frac{\Phi}{2}\right)^{-2}\right] .
\end{eqnarray} Another example is provided by a strongly interacting impurity
\cite{Castella_Zotos_1993,Matveev08}. In this case the number of expelled
particles $N$ is almost independent on the state of the impurity and may be
considered as a non-dynamic constant.  The remaining dependence of energy on
the superfluid phase $\Phi$ has a standard Josephson form
\begin{eqnarray}
  \label{eq:josephson} H_\mathrm{d}(\Phi) = -E_\mathrm{J} \cos\Phi.
\end{eqnarray} The Josephson energy $E_\mathrm{J} = \hbar n V_\mathrm{c}$ is
expressed through the corresponding critical velocity $V_\mathrm{c}$, which in
this case is much smaller than the sound velocity $c$.

Expressions (\ref{eq:lagrangian_impurity}) or (\ref{eq:hamiltonian_impurity})
provide the core energy of the locally equilibrium depletion cloud
as a function of  its slow variables  $\Phi$ and $N$.
This procedure  emphasizes the fact that
introduction of $\Phi$ and $N$ does not rely on the
semiclassical interpretation of the condensate wavefunction. In fact, they may
be defined even away from the semiclassical weakly interacting regime, 
where the phase of the condensate as well as its depletion are 
not well defined.

\section{Coupling to phonons}
\label{sec:phonon_coupling}

\begin{figure}[b]
  \centering
  \includegraphics[width=0.7\columnwidth]{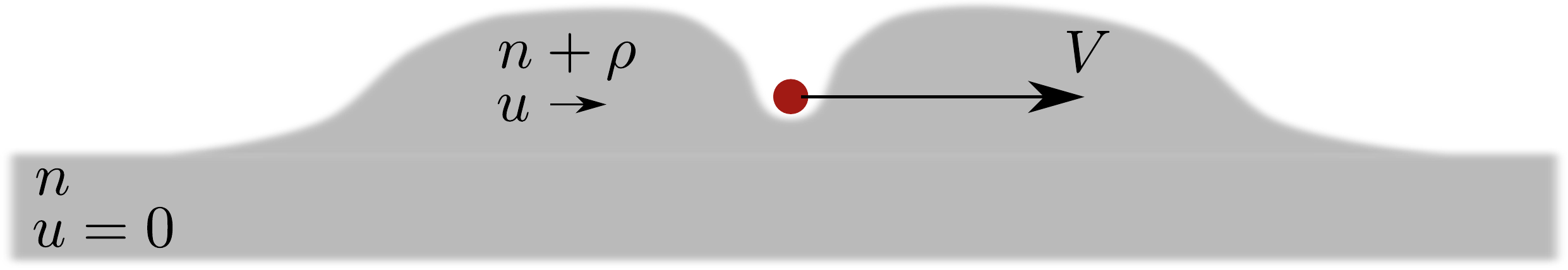}
  \caption{Impurity propagating in local environment.}
  \label{fig:local}
\end{figure}

An external force $F$ acting on the impurity
drives the system away from equilibrium,
making the impurity radiate energy and momentum.  For
a sufficiently weak force such a radiation takes the form of long wavelength
phonons, \emph{i.e.} small deviations of density $\rho(x,t)$ and velocity
$u(x,t)$ fields from their equilibrium values, see Fig.~\ref{fig:local}.
Below we show how coupling to phonons can be formulated in terms of the
collective variables $\Phi,N$. It turns out that this procedure is based
solely on the principles of gauge and Galilean invariance, \emph{i.e.} 
conservation of number of particles and momentum, and leads to universal 
results.

\subsection{Hydrodynamic description of phononic bath}
\label{sec:hydro}

We start by considering the Lagrangian, governing dynamics
of the phonon fields in the bulk of the liquid.
To this end it is convenient to parameterize them by
introducing the superfluid phase $\varphi(x,t)$
and the displacement field $\vartheta(x,t)$ such that
$u = \der_x \varphi/m$ and $\rho = \der_x \vartheta/\pi$.
The dynamics of these variables can be described
following the method of Popov \cite{PopovBookFunctional}
by considering slow change of the
density $n \to n+\rho(x,t)$ and, independently, the change
of the chemical potential,
\begin{eqnarray}
  \label{eq:chem_pot}
\mu \to  \mu - \dot\varphi(x,t) - \frac{mu^2(x,t)}{2}.
\end{eqnarray}
Substituting them into Eq.~(\ref{eq:grand-canonical_prime}) yields the Lagrangian of phonons,
\begin{eqnarray}
  \label{eq:lagrangian_bg} L_\mathrm{ph} = \int \!dx\Big[ p_0(\mu (x,t),
n(x,t))-p_0(\mu,n)\Big]= \int \!dx\left[-\rho\dot\varphi-
\frac{m(n+\rho)u^2}{2} -\Big(e_0(n+\rho)-e_0(n)-\mu\rho\Big)\right].
\end{eqnarray}

For nonzero phononic fields,
the impurity is subject to the modified local supercurrent and
chemical potential in the co-moving reference frame
\begin{eqnarray}
\label{eq:muprime}
\mu' &=& \mu -\dot\varphi - \frac{m u^2}{2} + \frac{m (V-u)^2}{2} =
\mu(n)+ \frac{mV^2}{2} -(\dot\varphi + V\derx\varphi)\,; \\
  \label{eq:jprime}
j' &=& -(n+\rho)(V-u) =  -nV -\frac{1}{\pi}
(\dot\vartheta +V\derx \vartheta)\,,
\end{eqnarray}
where the phonon variables are taken at the instantaneous
spatial position, $X(t)$  of the  impurity.
Equation~(\ref{eq:muprime}) follows from Eq.~(\ref{eq:chem_pot}) and the fact that
in the presence of the background flow $u=\derx\varphi/m$
the velocity of impurity with respect to the liquid is changed to $V-u$.
To derive expression (\ref{eq:jprime})
for the modified supercurrent we have used the
continuity equation in the form $\dot\vartheta/\pi  = -(n+\rho) u$. This
relation is an exact statement, which follows from the gauge invariance and
is valid for any configuration of the fields.

Substituting the modified supercurrent and chemical
potential, Eqs.~(\ref{eq:muprime}),~(\ref{eq:jprime}),
into Eq.~(\ref{eq:lagrangian_subst}) and
subtracting the corresponding equilibrium values,  results in the
following universal form of the interaction
Lagrangian,
\begin{eqnarray}
  \label{eq:imp-phonons} L_\mathrm{int} = \frac{1}{\pi}\,\Phi
\,\frac{\mathrm{d}}{\mathrm{d}t}\vartheta(X,t) +
N\frac{\mathrm{d}}{\mathrm{d}t}\varphi (X,t)  .
\end{eqnarray}
It is \emph{full time derivative}
${\mathrm{d}}/{\mathrm{d}t}= \der_t +\dot X\partial_x=\der_t +V\partial_x$ which enters
the interaction term, as follows from
Eqs.~(\ref{eq:muprime}),~(\ref{eq:jprime}).

The interaction Lagrangian, Eq.~(\ref{eq:imp-phonons}) provides dynamics of the
collective coordinates $N$ and $\Phi$.  It shows that the corresponding
canonical momenta are the phonon degrees of freedom at the location of the
impurity, i.e $\varphi(X,t)$ and $\vartheta(X,t)$ correspondingly. Through the
gradient terms in Eq.~(\ref{eq:lagrangian_bg}) these two local variables are
connected to the phonon fields elsewhere and it is the
dynamical properties of these phonons which determine the behavior of the
impurity.
For example if the spectrum of
phonons is discrete, one expects coherent oscillations of few modes. In the
infinite system the continuous spectrum of the background modes leads to
dissipation similar to that of Caldeira-Leggett model
\cite{Caldeira1983Quantum}.

\subsection{Linear phonons and transformation to chiral fields}
\label{sec:linear_chiral}

Away from this
interaction region the excitations of the liquid may be considered as linear
ones.  Therefore one may keep only the quadratic terms in the phononic
Lagrangian Eq.~(\ref{eq:lagrangian_bg}).
For the kinetic energy in Eq.~(\ref{eq:lagrangian_bg}) one thus retains
the leading term $mnu^2/2$, while the potential energy is expanded as
$e_0(n+\rho)-e_0(n)-\mu\rho\simeq (\der\mu/\der n)\rho^2/2 = (mc^2/2n)\rho^2$,
using thermodynamic relation between the compressibility and the sound
velocity $c$. As a result, one obtains the quadratic Luttinger liquid
Lagrangian
\begin{eqnarray}
   \label{eq:LLL}
L_\mathrm{ph} =
   \frac{1}{\pi}\int \!dx
 \left[ -\derx\vartheta\dert\varphi -\frac{c}{2K}
   (\derx \vartheta)^2
  -\frac{cK}{2}\left(\derx\varphi\right)^2 \right]\,.
\end{eqnarray}
Besides the sound velocity $c$ the Lagrangian in Eq.~(\ref{eq:LLL}) is
characterized by the dimensionless Luttinger parameter $K=\pi n/mc$, which
depends on the degree of correlations in the host liquid
\cite{HaldanePRL81}. For a liquid of weakly repulsive bosons the Luttinger
parameter $K$ is large, $K\gg 1$ but can be reduced down to the limiting value
$K=1$ by increasing the repulsive interactions or decreasing the density of
the particles \cite{Gangardt2003Stability}. 
The combination $\pi c/K=mc^2/n$ entering Eq.~(\ref{eq:LLL})
contains information about interactions between the particles in the liquid,
while the combination $c K/\pi=c\kappa = n/m$ is independent of the
interactions as a consequence of the Galilean invariance.

To gain an additional insight into the physics of the depleton--phonon
interaction,  the phononic fields can be  decomposed into a doublet of
right- and left-moving chiral components with the help of
the linear transformation
\begin{eqnarray}
\label{eq:phipm}
\begin{pmatrix}\vartheta/\pi \\\varphi\end{pmatrix} =\mathsf{T}  \chi
= \mathsf{T}
\begin{pmatrix}\chi_+ \\ \chi_- \end{pmatrix}
 \,,\qquad
\mathsf{T} =\frac{1}{\sqrt{2}}
\begin{pmatrix}\scriptstyle\sqrt{\kappa} & \scriptstyle\sqrt{\kappa}\\
\frac{1}{\sqrt{\kappa}} & - \frac{1}{\sqrt{\kappa}}    \end{pmatrix}\, ,
\end{eqnarray}
where $\kappa=K/\pi=n/mc$. 
In terms of the chiral fields
the Lagrangian, Eq.~(\ref{eq:LLL}) splits into a sum of two independent
contributions,
 \begin{eqnarray}
   \label{eq:LLLch}
L_\mathrm{ph}[\chi] =
   \frac{1}{2}\int \!dx
 \left[\chi_+ (\der_x\der_t + c\der_x^2)\chi_+ +
\chi_-(-\der_x\der_t+c\der_x^2)\chi_- \right]\, = \frac{1}{2}\!  \int\! dx\, \chi^\dagger
\mathsf{D}^{-1} \chi\,   .
 \end{eqnarray}
The matrix of inverse phonon propagator is defined by its
Fourier representation,
\begin{eqnarray}
  \label{eq:dinverse}
\delta(x)\delta(t)\mathsf{D}^{-1}
 =
\int\frac{\mathrm{d}q}{2\pi}\frac{\mathrm{d} \omega}{2\pi}\,
e^{iqx-i\omega t} \mathsf{D}^{-1} (q,\omega)
, \qquad
\mathsf{D}^{-1} (q,\omega) =
    \begin{pmatrix} q(\omega-cq) & 0\\ 0 & -q (\omega +cq)\end{pmatrix} \, .
\end{eqnarray}
The equation of motion following from the Lagrangian (\ref{eq:LLLch}) dictate
a simple coordinate and time dependence, $\chi_\pm(x,t) =
\chi_\pm(x\mp ct)$. Using this fact one can show that for uniformly moving
reference point $X=Vt$ one has   
\begin{eqnarray}
  \label{eq:dt_chiral}
 \frac{\mathrm{d}}{\mathrm{d}t} \chi_\pm (X,t)= 
 (V\mp c)\der_x \chi_\pm (X,t)\, .
\end{eqnarray}
This property is in fact
a statement about correlation functions of the fields $\chi_\pm$
calculated with the Gaussian action, Eq.~(\ref{eq:LLLch}) which enforces
classical equations of motion;
the path integration is performed over arbitrary configurations of the
fields.  
The interaction term, Eq.~(\ref{eq:imp-phonons})
may be conveniently rewritten by introducing chiral collective variables
\begin{eqnarray}
  \label{eq:yupsilon}
  \Lambda = \begin{pmatrix}\Lambda_+\\\Lambda_- \end{pmatrix}=
  \mathsf{T^\dagger}\begin{pmatrix}\Phi\\N \end{pmatrix}\, .
\end{eqnarray}
In the static limit the quantities $\Lambda_\pm$ are proportional, up to factor of
$\sqrt{2\pi}$,  to the
chiral phase  shifts $\delta_\pm$  introduced in
Refs.\cite{Imambekov08,Kamenev08}.
In presence of external force $F$  they acquire the dynamics,
which is governed by
the total Lagrangian of the depleton interacting with
the phonons
\begin{eqnarray}
  \label{eq:int_phonons_upsilon} L_\mathrm{tot} = P\dot X-
  H(P, \Lambda)+U(X) -\dot
\Lambda^\dagger(t)\,\chi(X,t) + L_\mathrm{ph}[\chi]
\end{eqnarray}
where the impurity Hamiltonian $H(P,\Lambda)$ is obtained from
Eq.~(\ref{eq:hamiltonian_nophonons}) by using the linear relation,
Eq.~(\ref{eq:yupsilon}) between the chiral phase shifts
$\Lambda$ and collective variables $\Phi$ and $N$, and $L_{\mathrm{ph}}$ is given by Eq.~(\ref{eq:LLLch}). Here $U(X)$ is an external potential acting on the impurity atom only and $F=-\partial U/\partial X$ is the external force.

\subsection{Integrating out the phonons}
\label{sec:int_phonons}

We have now all necessary ingredients for describing dynamics of impurity
coupled through the interaction term, Eq.~(\ref{eq:imp-phonons}) to the
phononic bath.  The presence of the impurity is felt by phonons through
time-dependent boundary conditions at $x=X(t)$ parameterized by the collective
variables $N(t)$ and $\Phi(t)$, or equivalently by chiral phases
 $\Lambda_\pm(t)$.  Here we
simplify even further our description by solving phononic linear equations of
motion for any variation of these collective variables and substituting the
obtained solution back into the action.  This procedure leads to dynamics of
the impurity expressed in terms of collective variables only and is equivalent
to exact integration of Gaussian phononic action.

To this end we employ the Keldysh formalism
\cite{Keldysh65,Kamenev_Levchenko_2009} and extend the dynamical variables
$X(t)$, $P(t)$, $\Lambda_\pm(t)$ as well as phononic fields $\chi_\pm(x,t)$ to
forward and backward parts of the closed time contour $t\to t_\pm$.
Performing Keldysh rotation, we write them as $X(t_\pm)=X_\mathrm{cl}\pm
X_\mathrm{q}/2$, $ \Lambda(t_\pm) = \Lambda_\mathrm{cl}\pm
\Lambda_\mathrm{q}/2$, and $\chi(t_\pm)=\chi_\mathrm{cl}\pm \chi_\mathrm{q}/2$
with the help of symmetric (``classical'', cl) and antisymmetric (``quantum'',
q) combinations.  The coupling term Eq.~(\ref{eq:imp-phonons}) then becomes
\begin{eqnarray}
  \label{eq:s_int_keldysh}
  L_\mathrm{int} =
  -\dot\Lambda^\dagger_\mathrm{cl} (t) \chi_\mathrm{q} (X_\mathrm{cl},t)
  -\dot\Lambda^\dagger_\mathrm{q}(t) \chi_\mathrm{cl} (X_\mathrm{cl},t)
  -\dot\Lambda^\dagger_\mathrm{cl}(t) X_q (t)\,\der_x\chi_\mathrm{cl}  (X_\mathrm{cl},t)
\end{eqnarray}
up to  terms linear in $X_\mathrm{q}$.
The advantage of  chiral fields introduced in  Eq.~(\ref{eq:phipm})
is that one can use the property~(\ref{eq:dt_chiral})
together with classical trajectory $X_\mathrm{cl} = V t$
to simplify the interactions, Eq.~(\ref{eq:s_int_keldysh}) as
\begin{eqnarray}
  \label{eq:s_int_keldysh_v}
  L_\mathrm{int} =
  -\dot\Lambda^\dagger_\mathrm{cl} \chi_\mathrm{q}
  +\left(\Lambda^\dagger_\mathrm{q}
  +X_\mathrm{q}\dot\Lambda^\dagger_\mathrm{cl} \mathsf{V^{-1}}\right)
   \mathrm{d}_t\chi_\mathrm{cl}
\end{eqnarray}
where we have introduced the matrix
\begin{eqnarray}
  \label{eq:vminusone}
  \mathsf{V}^{-1}=
\begin{pmatrix}  \frac{1}{c-V} & 0
\\  0  & -\frac{1}{c+V}\end{pmatrix}\, .
\end{eqnarray}

The interaction term, Eq.~(\ref{eq:s_int_keldysh_v})
is linear in phononic fields so that a Gaussian
integration with quadratic action, Eq.~(\ref{eq:int_phonons_upsilon})
can be performed by standard methods as explained in Appendix~\ref{sec:s_diss}.
It leads to quadratic, though  a nonlocal in time
effective action for the collective variables,
\begin{eqnarray}
  \label{eq:seff-lambda}
S_\mathrm{eff}
&=& -\frac{1}{2}\int\! \mathrm{d}t\; \dot
\Lambda^\dagger_\mathrm{cl} (t) \;
\left[\Lambda_\mathrm{q}(t) +\mathsf{V}^{-1}
    \dot \Lambda_\mathrm{cl}(t) X_\mathrm{q}(t)\right] \nonumber \\
    &-& \frac{1}{2}\int\! \mathrm{d}t \mathrm{d}t'
    \,\left[\Lambda^\dagger_\mathrm{q}(t)
      + X_\mathrm{q}(t)\dot\Lambda^\dagger_\mathrm{cl}(t)\mathsf{V}^{-1}\right]
    \partial_t\mathsf{F}(t-t')
    \left[\Lambda_\mathrm{q}(t') +
      \mathsf{V}^{-1}\dot\Lambda_\mathrm{cl}(t')X_\mathrm{q}(t')\right]\, ,
\end{eqnarray}
where, assuming thermal equilibrium of the phononic subsystem,
the matrix $\mathsf{F}(t)$ is related by inverse Fourier transform to
the matrix
\begin{equation}\label{eq:matrices}
 \mathsf{F} (\omega)= \begin{pmatrix}
    \coth\frac{\omega}{2T_+} & 0\\
    0 & \coth\frac{\omega}{2T_-}
    \end{pmatrix}\,,
\end{equation}
of thermal distribution of the chiral bosons
with the temperatures $T_\pm = T (1\mp V/c)$ modified by the corresponding
Doppler shifts.

One should supplement the action Eq.~(\ref{eq:seff-lambda})
with the Keldysh analogue of the action corresponding to the depleton
Hamiltonian, Eq.~(\ref{eq:hamiltonian_nophonons}),
\begin{equation}
\label{eq:action-impurity-Keldysh}
    S=\int dt
    \left[ \left( \dot X_\mathrm{cl}-\der_{P} H\right)P_\mathrm{q}
    -\left( \dot P_\mathrm{cl} -\partial_X U\right)X_\mathrm{q}
   - \nabla_{\Lambda} H\,\cdot
   \Lambda_\mathrm{q} \right]\,,
\end{equation}
where we kept only terms linear in the quantum components and
$H = H(P_\mathrm{cl},\Lambda_\mathrm{cl})$.
Notice that quadratic terms in quantum fields are absent in
Eq.~(\ref{eq:action-impurity-Keldysh})
while cubic and higher orders are omitted in the spirit
of the semiclassical approximation.

The second line in the effective action Eq.~(\ref{eq:seff-lambda})
may be split with the help of the
Hubbard-Stratonovich transformation, which introduces
two real uncorrelated Gaussian noises
$\xi_+(t),\xi_-(t)$. Their correlation matrix
in the frequency representation takes the standard Ohmic form
(see, \emph{e.g.} \cite{Caldeira1983Quantum}),
\begin{equation}
\label{eq:noise-correlator}
    \Big\langle \xi(\omega) \xi^\dagger(\omega)\Big\rangle
    =\omega \mathsf{F}(\omega)\,,\qquad
    \xi =\begin{pmatrix}\xi_+ \\\xi_-\end{pmatrix} 
\end{equation}
The action,
Eq.~(\ref{eq:seff-lambda}) becomes  local in time,
 \begin{eqnarray}
  \label{eq:seff-HStransformed}
S_\mathrm{eff}=- \frac{1}{2}\int\! \mathrm{d}t\,  
\left(\dot\Lambda^\dagger_\mathrm{cl} - 2\xi^\dagger
\right)
    \left(\Lambda_\mathrm{q} +
      \mathsf{V}^{-1}\dot\Lambda_\mathrm{cl}X_\mathrm{q}\right).
\end{eqnarray}
Now the entire semiclassical action is linear in quantum components and
integration over them enforces the delta-functions of the equation of
motions. While Eq.~(\ref{eq:velocity_phi_n}) remains intact, due to the
absence of $P_\mathrm{q}$ in the effective action
(\ref{eq:seff-HStransformed}), Eqs.~(\ref{eq:p_0}),~(\ref{eq:phi_0}) and
(\ref{eq:n_0}) are modified by the phonons:
\begin{eqnarray}
\label{eq:momentum_diss-Lambda}
  \dot P &=& F -
  \frac{1}{2}\dot\Lambda^\dagger\mathsf{V}^{-1}\dot\Lambda +
  \xi^\dagger \mathsf{V}^{-1}\dot\Lambda \,,
  \\
    \label{eq:eq_of_motion_Lambda}
  \frac{1}{2}\dot \Lambda &=&  -\nabla_\Lambda H +\xi\,,
\end{eqnarray}
where we have dropped subscripts for clarity.  The obtained equations include
additional dissipative terms involving time derivatives of the
collective variables $\Lambda$. They also include fluctuations coming from the
pair of Gaussian noises $\xi_\pm(t)$ correlated accordingly to  
Eq.~(\ref{eq:noise-correlator}).

\section{Depleton dynamics at zero temperature. Radiative corrections}
\label{sec:coupling_phonons}

Our goal is to discuss non-equilibrium solutions of the equations of motion in
the presence of a constant external force $F$.
Neglecting for a moment fluctuation terms and using transformation
Eq.~(\ref{eq:yupsilon}) the equations of motion,
Eqs.~(\ref{eq:momentum_diss-Lambda}),~(\ref{eq:eq_of_motion_Lambda})
can be rewritten in terms of the
collective variables $\Phi$ and $N$ as follows,
\begin{eqnarray}
\label{eq:momentum_diss}
  \dot P &=& F
  - \frac{1}{2}
  \left(\dot\Phi,\dot N\right) \mathsf{T}\mathsf{V}^{-1}\mathsf{T}^\dagger
  \begin{pmatrix}
      \dot\Phi \\ \dot N\end{pmatrix}
 =F
  - \frac{c}{c^2-V^2}
  \left( \frac{\kappa V}{2c}\dot \Phi^2 +
   \dot \Phi\dot N+\frac{V}{2\kappa c}\dot N^2\right)
  \\
    \label{eq:eq_of_motion_Phi}
  \frac{\kappa \dot \Phi}{2}  &=&
  -\partial_\Phi H
=    nV-\partial_\Phi H_\mathrm{d}
\\
  \label{eq:eq_of_motion_N}
    \frac{\dot N}{2\kappa} &=&
    -\partial_N H
=   -mV^2/2   -\mu(n) -\partial_N H_\mathrm{d}
\, .
\end{eqnarray}

The rate of energy radiated by phonons is obtained by
taking derivative with respect to time of
the total impurity energy,
\begin{eqnarray}
  W =\dot H-F\dot{X} = V(\dot P-F)-\dot \Lambda^\dagger \cdot \nabla_\Lambda H
  =V(\dot P-F)+\dot\Phi\partial_\Phi H  + \dot N\partial_N H .
  \label{eq:energy_diss_1}
\end{eqnarray}
Using Eq.~(\ref{eq:velocity_phi_n}) and equations of motion  either in the
form (\ref{eq:momentum_diss-Lambda}--\ref{eq:eq_of_motion_Lambda}) or
(\ref{eq:momentum_diss}--\ref{eq:eq_of_motion_N})
we obtain
\begin{eqnarray}
  \label{eq:energy_diss}
  W = -\frac{1}{2}
  \dot \Lambda^\dagger\left[\mathsf{1}+V\,\mathsf{V}^{-1}\right] \dot \Lambda 
  = -\frac{c^2}{c^2-V^2} \left( \frac{\kappa}{2}\dot\Phi^2  +
   \frac{V}{c}\dot \Phi\dot N+\frac{1}{2\kappa}\dot N^2
     \right)\,,
\end{eqnarray}
The dissipation of momentum, Eq.~(\ref{eq:momentum_diss}) and energy,
Eq.~(\ref{eq:energy_diss}) is a generalization of Eqs.~(22),(23) in
Ref.\cite{Pelinovsky1996Instabilityinduced} (up to a factor of two), where they were derived in the
context of dynamics of grey solitons.

According to Eqs.~(\ref{eq:p_0})--(\ref{eq:n_0}), in the absence of the external force $F=0$ there is a family of stationary
solutions of the equations of motion, which are characterized by a constant
velocity $V$ below some critical velocity $V_\mathrm{c}$.  This solutions
describe the dissipationless motion of the impurity consistent with the
superfluidity. Indeed, by neglecting the fluctuation terms we effectively put
the temperature to be zero, thus making the one-dimensional liquid superfluid.

On the first glance one can just solve the set of the evolutionary equations
(\ref{eq:velocity_phi_n}), (\ref{eq:momentum_diss})--(\ref{eq:eq_of_motion_N})
to fully describe the impurities dynamics. One needs to be careful, though,
because Eqs.~(\ref{eq:eq_of_motion_Phi}), (\ref{eq:eq_of_motion_N}) correspond
to the motion in the vicinity of the \emph{maximum} of the Hamiltonian $H$ and
therefore exhibit runaway instability.  The further look at this instability
shows that its characteristic rate is of the order $mc^2\sim \mu$, 
which is well outside the frequency range of applicability of the 
theory developed above.
In fact this high-frequency instability of $\Phi$ and $N$ evolution is a
direct analog of the well-known spurious self-acceleration of charges due to
the back reaction of the electromagnetic field \cite{*[{}] [{; Chapter~75}]
  Landau_Lifshitz_2_Classical_Theory_of_Fields}.  The recipe to overcome it
is, of course, also well-known: instead of trying to solve
equations of motion directly, one should perturbatively find how radiation
corrections influence the dynamics \cite{*[{}] [{; Chapter~2}]
  Ginzburg_Applications_of_Electrodynamics}.  
This strategy offers a convenient
analytical approach to treat the dynamics described by
Eqs.~(\ref{eq:velocity_phi_n}) and
(\ref{eq:momentum_diss})--(\ref{eq:eq_of_motion_N}). Below we apply it to
study modifications to Bloch oscillations which arise due to the phonon
radiation.

\subsection{Radiative corrections to
Bloch oscillations}
\label{sec:pert. theory}
At zero temperature with a sufficiently small applied force one expects
the system to adiabatically stay in the ground state with total
momentum $P$, while the latter is changed by the external force $P=Ft$. 
In this zeroth approximation, the motion is nothing but a
tracing of the dispersion relation $E(P,n)$ and the phononic subsystem gains
no share of the work done on the system by $F$. The velocity is simply
\begin{equation}
V^{(0)}(t) =V_0(Ft,n) = \partial E(P,n)/\partial P\Big|_{P=Ft}\,.
\end{equation}
Since the dispersion relation displays a periodic behavior, one immediately
obtains velocity Bloch oscillations with period $\tau_\mathrm{B}^{(0)}=2\pi n/F$,
amplitude $V_\mathrm{c}$ and zero drift. 
In reality, the slow acceleration of the
impurity over the course of a Bloch cycle gives rise to a soft radiation of
low energy phonons, which serve to renormalize the period, amplitude and drift
from the zeroth order approximations.

To study the corrections to the depleton trajectory, let us assume it exhibits
a steady-state motion such that $V(t+\tau_\mathrm{B})=V(t)$, 
$N(t+\tau_\mathrm{B})=N(t)$ and
$\Phi(t+\tau_\mathrm{B})=\Phi(t)+2\pi$. 
Then it follows from Eq.~(\ref{eq:momentum})
that $P(t+\tau_\mathrm{B})=P(t)+2\pi n$.  Here $\tau_\mathrm{B}$ 
is the, \emph{a priori}
unknown, true period of the motion, not to be confused with the zeroth
approximation $\tau_\mathrm{B}^{(0)}$.
To find it we integrate Eq.~(\ref{eq:momentum_diss}) over a single Bloch cycle
\begin{equation}
  \label{eq:tau0}
  2\pi n=\int_0^{\tau_\mathrm{B}} dt \dot{P}=F\tau_\mathrm{B}-
  \frac{1}{2}\int_0^{\tau_\mathrm{B}} 
  dt\frac{1}{c^2-V^2}
  \left( \kappa V \dot \Phi^2 +
    2c\dot \Phi\dot N+\kappa^{-1}V\dot N^2\right) = \tau_\mathrm{B} (F-F_\mathrm{rad})
\end{equation}
where $F_{\textrm{rad}}$ is the average radiative friction
force exerted on the impurity over a single Bloch cycle.
Since the radiative frictional force tends to reduce the applied
force, Eq.~(\ref{eq:tau0}) indicates that the true period of oscillation
 is \emph{larger} than the zeroth approximation $\tau_\mathrm{B}^{(0)}$.

The work of the external force per unit time is given by
\begin{equation}
\label{eq:total energy}
FV=\dot{E}(P,n)-W.
\end{equation}
The first term in the r.h.s. of this equation is the reversible
change in energy of the
impurity, while the second term, owing to Eq.~(\ref{eq:energy_diss}), is the rate of energy channeled into the phonon system. We average
Eq.~(\ref{eq:total energy}) over a single Bloch cycle,  noticing that
$\langle\dot{E}\rangle=0$  due to the
periodicity of the dispersion relation.  The remaining term corresponds to the
power radiated into phononic
bath and leads to the \emph{drift velocity}:
\begin{equation}
\label{eq:V drift}
V_\mathrm{D}=-\langle W \rangle/{F}.
\end{equation}
Assuming  the energy pumped into the phonon system
per Bloch cycle to be small we use the bare trajectories.
$V_0(P,n)$, $\Phi_0(P,n)$, $N_0(P,n)$.  Using  the fact that
$dt=dP/F$ one can show that
\begin{equation}
\label{eq:V drift2}
V_\mathrm{D}=\sigma F.
\end{equation}
Here $\sigma$ is the $T=0$ mobility of the impurity,
given  by the average over the Brillouin zone
\begin{equation}
  \label{eq:mob}
  \sigma=\frac{1}{2\pi n}\int\limits_{-\pi n}^{\pi n}dP
  \left(\frac{c^2}{c^{2}-V_0^{2}}\right)
  \left[\frac{\kappa}{2}
    \left(\frac{\partial\Phi_{0}}{\partial P}\right)^{2}+
    \frac{V_0}{c}\left(\frac{\partial N_{0}}{\partial P}\right)
    \left(\frac{\partial\Phi_{0}}{\partial P}\right)
    +\frac{1}{2\kappa}\left(\frac{\partial N_{0}}{\partial P}\right)^{2}\right].
\end{equation}
It was mentioned in Sec.~\ref{sec:dyn_eq} that the equilibrium functions
$\Phi_0(P,n)$ and $N_0(P,n)$ can be obtained directly from partial derivatives
of the equilibrium dispersion relation $E(P,n)$. Since $V_0=\partial
E/\partial P$, the mobility may be expressed entirely in terms of $E(P,n)$ and
the Luttinger parameter $K=\pi\kappa$.

The fact that the mobility  can be  expressed
through equilibrium properties is reminiscent to the Kubo linear response
formulation. It is crucial to mention that they are {\em
  not} a linear response property in the Kubo sense. The Kubo linear response
takes place at finite  temperature, see Section \ref{sec:noise}.
At $T=0$ the liquid is superfluid and the impurity undergoes Bloch
oscillations with the amplitude $V_c$ at arbitrarily small external force
$F$. It means that the response is essentially non-linear. The mobility
$\sigma$ describes the average (over one period) shift of the oscillation center
due to the energy radiated in the course of such {\em non-linear}
oscillations. The fact that it may be fully expressed through the equilibrium
properties is rather remarkable on its own right.

The result in Eq.~(\ref{eq:V drift2}) holds for sufficiently weak external
perturbation $F<F_\mathrm{max}$.
One can estimate $F_{\max}$  by comparing the corresponding drift velocity
with the velocity of sound. From Eq.~(\ref{eq:V drift2}) we have
\begin{equation}
\label{eq:F_max}
 F_{\textrm{max}} = c/\sigma.
\end{equation}
As the force increases the separation of
length and energy scales used to define the depleton dynamics
cannot be justified. In other words for a strongly perturbed system
the \emph{equilibrium} dispersion relation $E(P,n)$ ceases to be a meaningful
concept.

Below we use  a model of an impurity  coupled via
delta-function interaction with
strength $G$ to the background particles to illustrate
the dynamical properties of a depleton discussed above.
We discuss two regimes: the strong  coupling regime,
where the interactions in the liquid can be of arbitrary strength and the weak coupling regime where the dynamics of the depletion cloud
is governed by Gross-Pitaevskii equation. In the latter case one is restricted
to a weakly interacting bosonic background.

\subsubsection{Strong coupling regime $G/c\gg 1$}
\label{sec:strong}

The impurity expels a large number $N\gg 1$ of particles from its vicinity
thus one can neglect the dynamics of $N$ and use the Josephson form,
Eq.~(\ref{eq:josephson}) for the remaining dependence of energy on superfluid
phase.  Due to the fact that velocity is bound by the critical
value $V_\mathrm{c}$ the impurity is slow $V\le V_\mathrm{c} \ll c$.
Indeed, the calculation for weakly interacting bosonic background,
Eq.~(\ref{eq:V N for large G}) (see also Ref.\cite{Gunn_Taras_PhysRevB.60.13139} ) gives $V_c=c^2/2G\ll c$.  Therefore,
if the impurity is not too heavy, $M\sim m$, we can neglect the second term in
the momentum, Eq.~(\ref{eq:momentum}) and we are left with the superfluid
contribution $P=n\Phi$ only.  Using this fact in Eq.~(\ref{eq:mob}) gives the
universal result for the mobility
\begin{equation}
\label{eq:mob_large_G}
\sigma=\frac{\kappa}{2\hbar n^2}=
\frac{K}{h n^2}=\frac{1}{2nmc}\, ,
\end{equation}
where we restored $\hbar=h/2\pi$. Note that in the strong coupling limit,
the mobility is \emph{independent} of
the impurity parameters  and \emph{only}
depends on the parameters of the host liquid, namely the Luttinger parameter
$K$ and the asymptotic density $n$. This result has been obtained by Castro-Neto
and Fisher, \cite{castro96} by using  the linear response approach.  

For the case of impenetrable bosons or free fermions corresponding
to $K=1$ one can use the analogy from the electronic transport.
Suppose the background is made of non-interacting fermions each
carrying electric charge $e$. In the frame co-moving with the
impurity a current $I=enV_{\mathrm{D}}$  flows through the wire, whose
quantum resistance is $R$. The latter is given by the Landauer
formula, which for spinless case reads as $R= h/e^2$.  The ohmic
power transferred to the system, $I^2 R$, must be supplied by the
external force, $FV_\mathrm{D}=h n^2V_\mathrm{D}^2$ giving the
result in Eq.~(\ref{eq:mob_large_G}) for $K=1$. For $K>1$ one
recalls that $R=h/K e^2$ \cite{Kane_Fisher_PhysRevB.46.15233}  leading again to
Eq.~(\ref{eq:mob_large_G}). Notice that discussion of
Ref.~\cite{Maslov_Stone_PhysRevB.52.R5539} claiming interaction-independent
mobility   is not applicable here, 
since we always assume that the system length is much larger than the
characteristic wavelength of phonons.

Scaling with $K$ implies that the mobility is higher for weaker interactions
and diverges in the free-boson limit. To understand this result, notice that
the moving impenetrable impurity experiences $nV_\mathrm{D}$ collisions per
unit time.  A given collision  results in the momentum
transfer $m V_{\mathrm{D}}$ to the boson of the gas.
The balance of forces $F=n m  V_\mathrm{D}^2$
then implies  $V_\mathrm{D}=F/\sqrt{n m   F}$ leading to
$ \sigma = 1/\sqrt{n m   F}\to\infty$ as
$F\to 0$, in agreement with non-interacting limit 
of Eq.~(\ref{eq:mob_large_G}).

\begin{figure}[t]
\centering
\includegraphics[width=0.6\columnwidth]{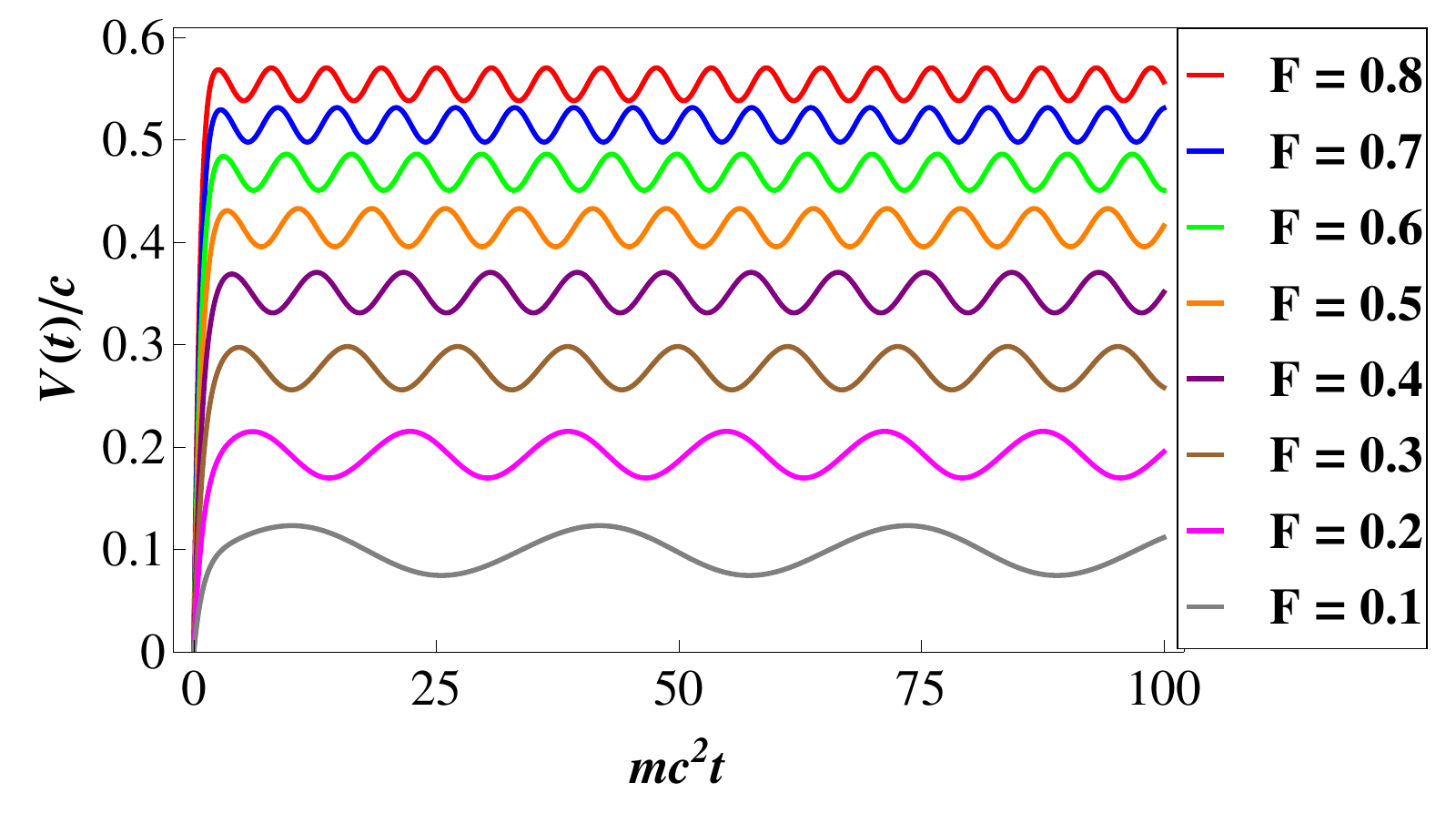}
\caption{Velocity as a function of time for various forces listed in the
  legend ($F$ in units of $F_{\mathrm{max}}=2nmc^{2}$). Here $\kappa=20$, $G/c=20$ and
  $M=40m$. The dashed lines correspond to the drift velocity plotted in
  Fig.~\ref{fig:drift_vs_F}. One notices that as $F$ increases, the drift
  velocity and frequency of oscillations increases, while the velocity
  amplitude (as measured from $V_{\mathrm{D}}$) decreases.}
\label{fig:V_vs_t}
\end{figure}

Turning to the period of oscillations, Eq.~(\ref{eq:tau0})
we use the relations $V\approx V_{\mathrm{D}}=\kappa F/2n^2$,
$\dot{\Phi}\approx F/n$ to obtain the
renormalized period of Bloch oscillations,
\begin{equation}
\label{eq:tau}
\tau_\mathrm{B}\approx\frac{2\pi n}{F}\left[1+\left(\frac{F}{2nmc^2}\right)^2
\right].
\end{equation}
The characteristic force entering Eq.~(\ref{eq:tau})
coincides with the  upper bound $F_\mathrm{max}=2nmc^2$ obtained from
Eq.~(\ref{eq:F_max}) using the result (\ref{eq:mob_large_G}) for the mobility.
Indeed, the frequency of the motion $\omega\approx
F/n$ should be small compared to the typical phonon frequency $mc^2$ in order
to justify the large scale separation employed in
Sec.~\ref{sec:qualiative-analysis}.

To go beyond lowest order in $F$ we devise a numerical approach to the
strongly coupled impurity based on the Josephson form,
Eq.~(\ref{eq:josephson}) for the energy. Using it in
Eq.~(\ref{eq:eq_of_motion_Phi}) and recalling Eq.~(\ref{eq:momentum_diss})
gives,
\begin{equation}
\dot{P}=F-\frac{1}{2}\frac{\kappa V}{c^2-V^2}\dot{\Phi}^2;\,\,\,\,\,\,\quad\quad \frac{\kappa\dot{\Phi}}{2n}=V-V_{c}\,\textrm{sin}\Phi,
\label{eq:eq. of motion 2}
\end{equation}
where $P$ is given by Eq.~(\ref{eq:momentum}). The second of
Eqs.~(\ref{eq:eq. of motion 2}) relates the change in $\Phi$ with the
deviation of the impurity velocity $V$ from its equilibrium value
$V_c\sin\Phi$. The velocity $V(t)$ is obtained by solving
Eqs.~(\ref{eq:eq. of motion 2}) numerically and the results
depicted in Figs.~\ref{fig:V_vs_t} agree well with the ansatz
(\ref{eq:velocity_vs_t}). As one can see the precise choice of
initial conditions is essentially irrelevant to the ensuing discussion, where
we focus only on the asymptotic, steady state behavior of the functions
$\Phi(t)$ and $V(t)$.

After a sufficiently long time  the  system
reaches the regime  of steady Bloch oscillations, where its velocity
is given by 
\begin{equation}
\label{eq:velocity_vs_t}
V(t)=V_\mathrm{D}+V_\mathrm{B}\cos(\omega_\mathrm{B}t + \delta)\, .
\end{equation}
Here the parameters $V_\mathrm{D}$, $V_\mathrm{B}$,
$\omega_\mathrm{B}=2\pi/\tau_\mathrm{B}$ and  $\delta$
depend on  the ratio of the external force $F$ to the critical force
$F_\mathrm{max}$.

\begin{figure}[t]
\centering
\includegraphics[width=0.47\columnwidth]{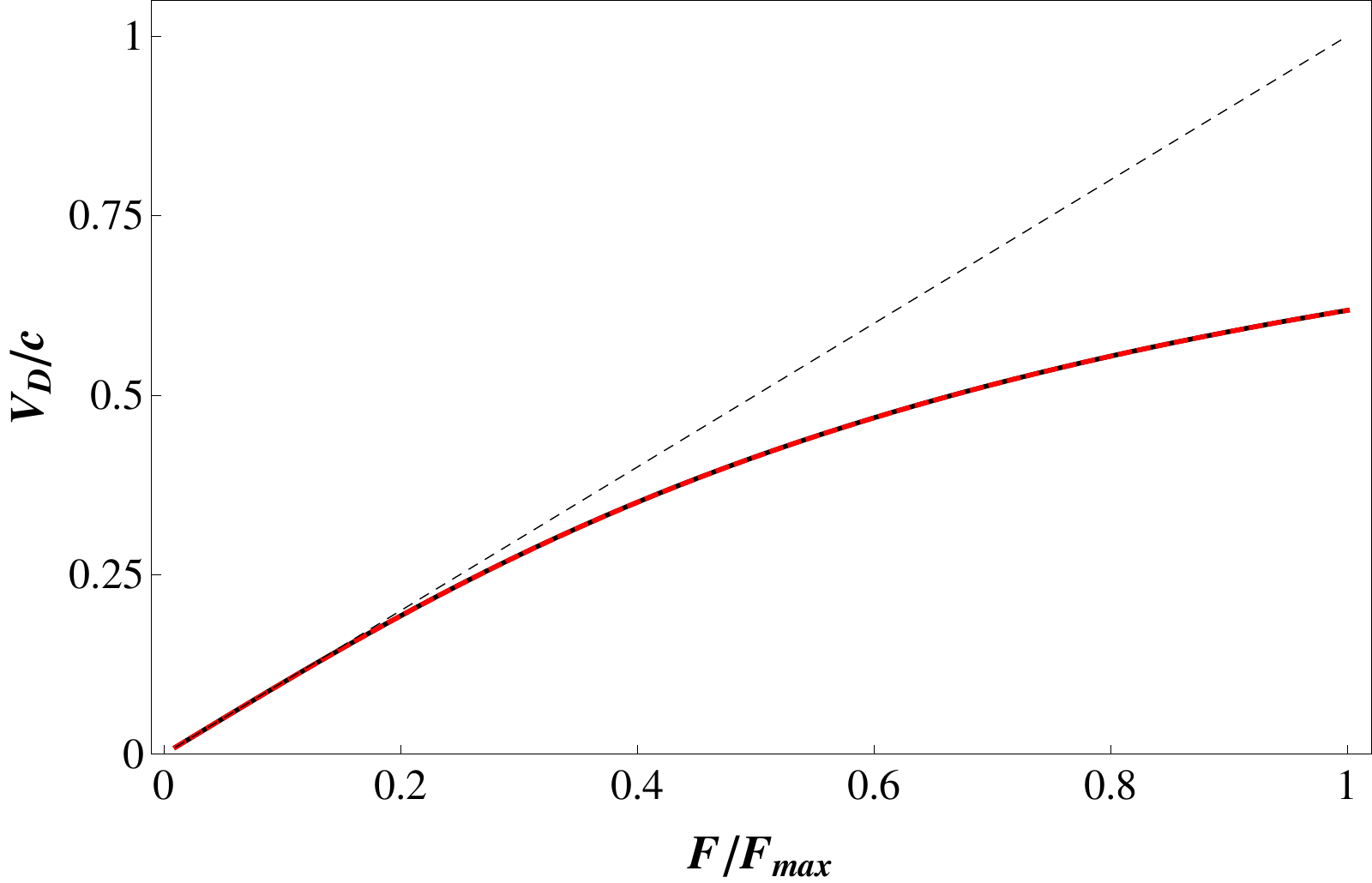}\hspace{0.7cm}
\includegraphics[width=0.48\columnwidth]{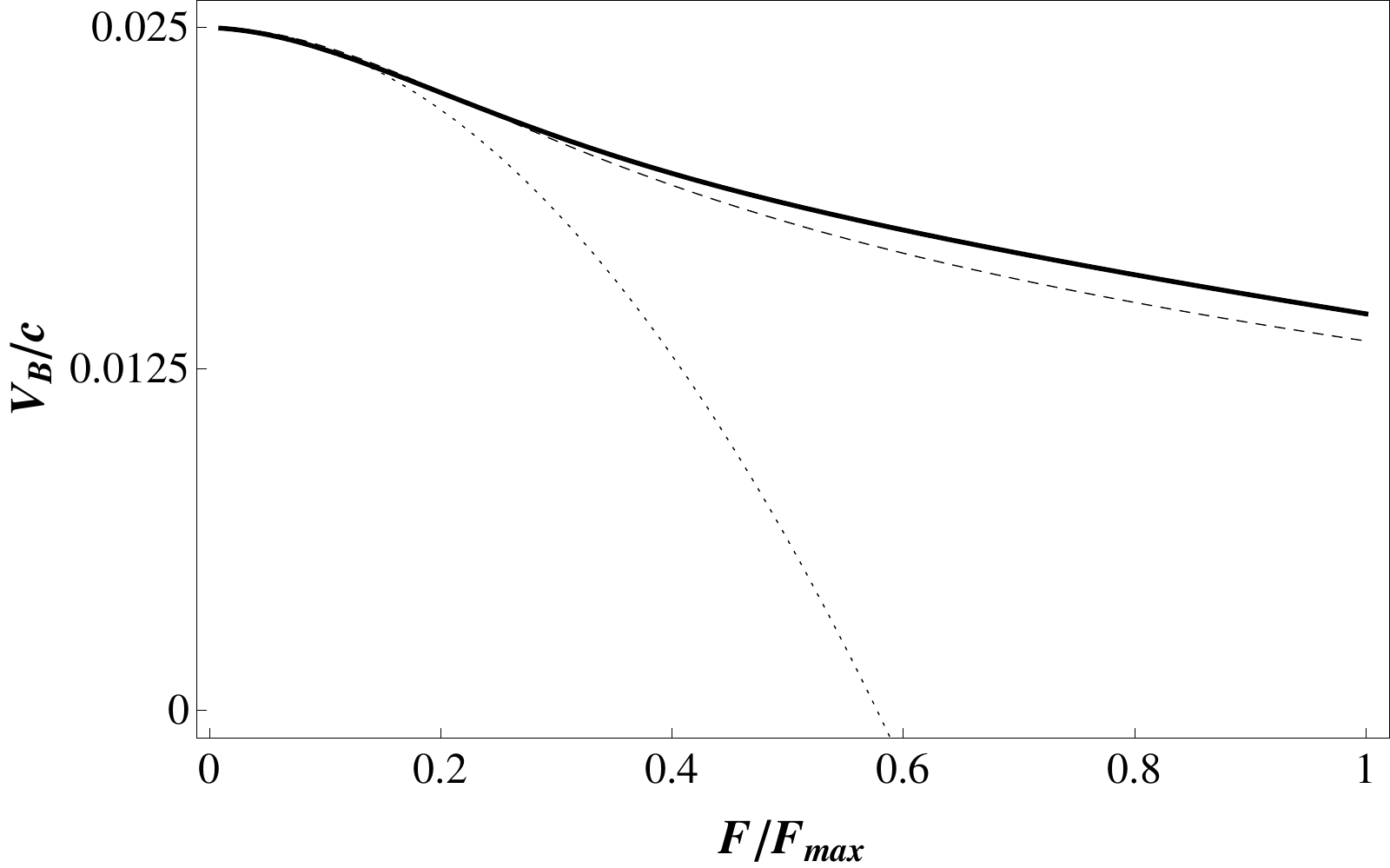}
\caption{ Drift velocity $V_\mathrm{D}$ (left panel) and  
  amplitude $V_\mathrm{B}$ (right panel)  as
  functions of the applied force $F$ corresponding to the same set of
  parameters are as in
  Fig.~\ref{fig:V_vs_t}. 
  The solid curve on left panel is indistinguishable from the analytic
  prediction, the first of Eqs.~(\ref{eq:drift_and_amp}), while 
  the dashed curve is the small force result Eq.~(\ref{eq:V drift2})
  with mobility $\sigma=\kappa/2\hbar n^2$.  On right panel  the solid curve
  is numerics, the dashed curve is Eq.~(\ref{eq:VB_theory}),
  and the dotted curve corresponds to the small F limit of Eq.~(\ref{eq:vb}).}
\label{fig:drift_vs_F}
\end{figure}

To find the drift velocity and period of Bloch oscillations 
we substitute the ansatz $V = V_\mathrm{D}$ and $\dot P =n \dot \Phi =
\omega_\mathrm{B}$  into Eqs.~(\ref{eq:eq. of motion 2}) to obtain 
a closed set of equation, 
\begin{equation}
  n\omega_\mathrm{B}=F-\frac{1}{2}\frac{\kappa V_\mathrm{D}\omega_\mathrm{B}^{2}}
  {c^{2}-V_\mathrm{D}^{2}};\qquad
  \frac{\omega_\mathrm{B}}{2mc^{2}}=V_\mathrm{D}/c
  \label{eq:3}
\end{equation}
which are solved to give
\begin{equation}
  \label{eq:drift_and_amp}
  \frac{V_\mathrm{D}}{c} =\frac{\omega_\mathrm{B}}{2mc^2} =
  \frac{F_\mathrm{max}}{2 F}
    \left(
    \sqrt{1+\left(\frac{2F}{F_\mathrm{max}}\right)^2}-1\right)\, .
\end{equation}
Expanding the expression for
$\omega_\mathrm{B}=2\pi/\tau_\mathrm{B}$
for small $F/F_\mathrm{max}$ one recovers the result (\ref{eq:tau}) obtained
by a different method.
The results for the drift velocity are plotted in 
Fig.~\ref{fig:drift_vs_F}a and are in excellent agreement with numerical
solution of Eqs.~(\ref{eq:eq. of motion 2}). 

The  Bloch amplitude $V_\mathrm{B}$ can also be estimated from
Eqs.~(\ref{eq:eq. of motion 2}).  In Appendix~\ref{sec:solution_strong} 
it is shown that
\begin{eqnarray}
  \label{eq:vb}
  \qquad V_\mathrm{B} =  V_\mathrm{c}
  \left\{1-\left[1+\frac{1}{2}
      \left(\frac{M-mN}{m\kappa}\right)^2\right]\left(\frac{F}
    {F_{\mathrm{max}}}\right)^2\right\}\, .
\end{eqnarray}
Thus, Eq.~(\ref{eq:vb}) predicts a decrease of the Bloch amplitude $V_{\mathrm{B}}$ with
increasing $F$ (see Fig.~\ref{fig:drift_vs_F}b). Physically sensible, this
result implies that as the force increases, the ideal equilibrium tracing of
the dispersion relation is, to a degree, lost. This is due to the phononic
subsystem gaining a proportionately larger share of the work done by the
external force.  Expression (\ref{eq:vb}) is compared with results of the
numerical solution of Eqs.~(\ref{eq:eq. of motion 2}) and the results
presented in the right panel of Fig.~\ref{fig:drift_vs_F}b show a good
agreement.

\subsubsection{Weakly interacting background bosons and
weak coupling regime $G/c\ll 1$}
\label{sec:weak}

\begin{figure}[b]
\centering
\includegraphics[width=0.6\columnwidth]{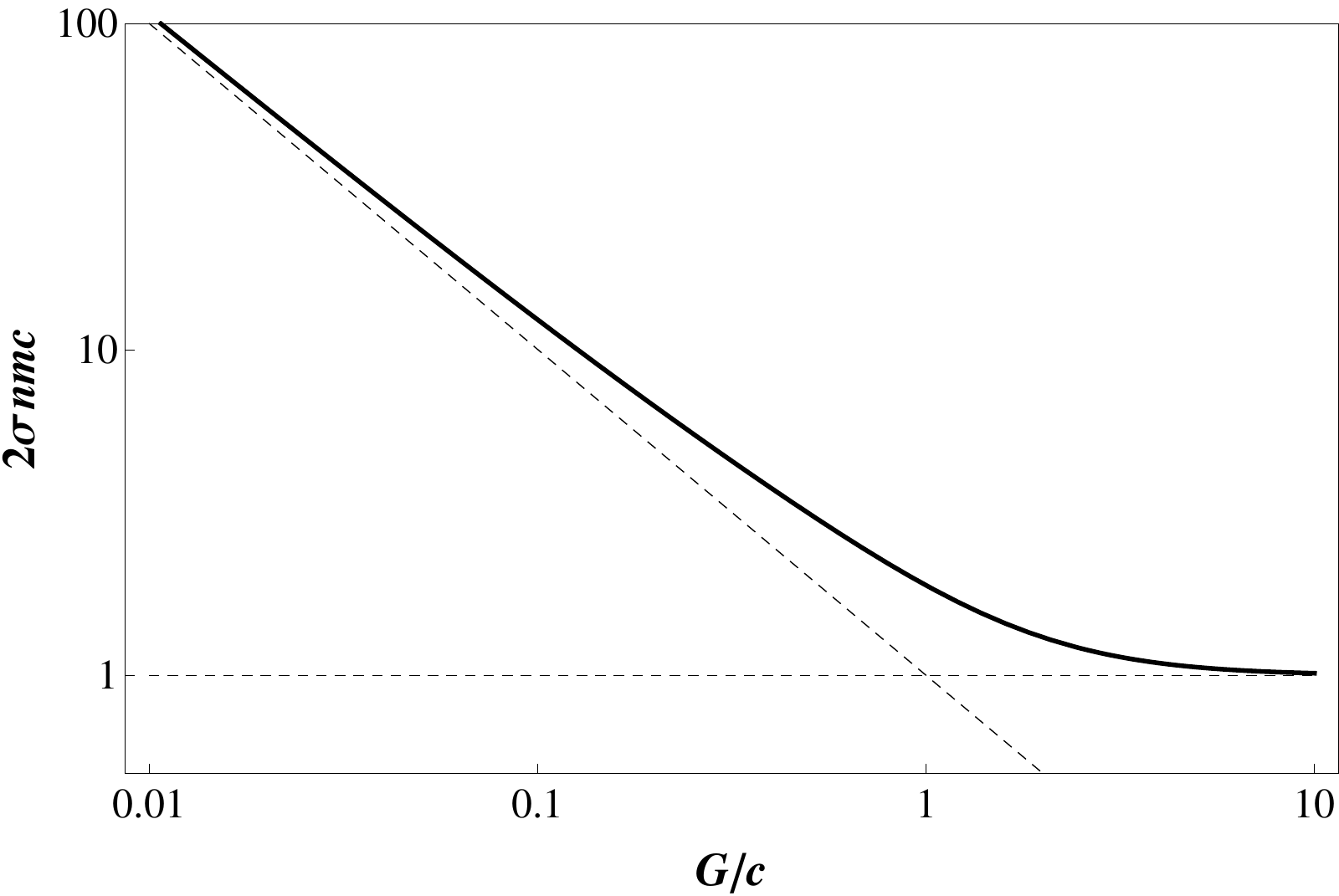}
\caption{Log-log plot of the \emph{zero temperature} mobility $\sigma$ as a
function of the impurity coupling $G$ in a weakly interacting superfluid
(solid line). The dashed lines correspond to the asymptotic values of $\sigma$
in the strong and weak coupling limits given by Eqs.~(\ref{eq:mob_large_G}) and (\ref{eq:mob_small_G}), respectively (with $\hbar=1$). The Luttinger parameter for the liquid is taken to be $\kappa=n/mc=20$, while $M=m$.}
\label{fig:mobility}
\end{figure}

We deal with the case of the background made of
weakly interacting bosons by using the Gross-Pitaevskii
equation for the background Bose liquid as explained in
Appendix~\ref{sec:equilibrium impurity}.
By calculating numerically
the dispersion relation of the impurity, 
see \emph{e.g.}  Fig.~\ref{fig:dispersion},
we obtain  the functions $N_0$ and $\Phi_0$ for all values of
the impurity-background coupling $G$. The results of
numerical evaluation of Eq. (\ref{eq:mob}) are shown in Fig.~\ref{fig:mobility}.


Let us turn now to the case of weak coupling $G\ll c$,
where particles can tunnel through the semi-transparent impurity.
In such a case one expects the mobility to increase
as the impurity becomes more transparent.
For weak coupling, the main contribution to the integral in Eq.~(\ref{eq:mob})
comes from the region where the velocity of the impurity is maximal, \emph{i.e.}
$V\sim V_\mathrm{c}$.
It corresponds to the inflection points of the dispersion,
Fig.~\ref{fig:dispersion}. Beyond this point the dispersion follows closely
the dispersion of grey solitons, Eq.~(\ref{eq:ds_characteristics})
and one can use the expressions (\ref{eq:momentum_energy_soliton})
to estimate the mobility.  As the momentum $P_\mathrm{c}$
corresponding to maximum velocity $V(P_\mathrm{c}) = V_\mathrm{c}$ is small we
simplify
Eqs.~(\ref{eq:ds_characteristics}),~(\ref{eq:momentum_energy_soliton}) and
write
\begin{eqnarray}
  \label{eq:vphismall}
  \frac{V(\Phi)}{c}\simeq 1-\frac{1}{2}\left(\frac{\Phi}{2}\right)^2;\qquad
  \frac{P(\Phi)}{n}=\Phi-\sin\Phi\simeq \frac{\Phi^3}{6}\,
\end{eqnarray}
or, equivalently,
\begin{eqnarray}
  \label{eq:vpsmall}
  \Phi(P) = (6P/n)^{1/3};\qquad \frac{V(P)}{c} = 1-\frac{(6P/n)^{2/3}}{8}\, .
\end{eqnarray}
Substituting these expressions into Eq.~(\ref{eq:mob}) we can estimate the
mobility as
\begin{equation}
  \label{eq:mob_small_G}
  \sigma\sim \frac{\kappa }{n} \int_{P_\mathrm{c}} \frac{dP}{1-V^2/c^2}
  \left(\frac{\der \Phi}{\der P}\right)^2 \sim \frac{\kappa}{n^2}
  \int_\Lambda \frac{d P/n}{(P/n)^2}\sim \frac{\kappa}{n^2}\frac{c}{G}=
  1/nmG\, .
\end{equation}
Here we have estimated the momentum cutoff $\Lambda =P_\mathrm{c}/n\sim G/c$
using Eq.~(\ref{eq:vc}) for the critical velocity and the second equation in
(\ref{eq:vpsmall}).  The coefficient in Eq.~(\ref{eq:mob_small_G}) is fixed
from the numerics and was found to be very close to one for $M=m$.
Interestingly, in the weak coupling limit $\sigma$ is \emph{independent} of
the correlations within the liquid (provided the liquid is weakly interacting,
$G\ll c$), diverging in the limit of a completely transparent impurity. Thus,
in the weak coupling limit, the upper critical force $F_\mathrm{max} =nGmc$
corresponds to energy difference $Gn$ per healing length $\xi=1/mc$.

\section{Finite temperature dynamics of depleton. 
Backscattering of thermal phonons 
and viscous friction}
\label{sec:noise}

At a finite temperature, even in the absence of the external force,
the collective coordinates fluctuate
around their equilibrium position $\Lambda_\pm^{(0)}$, found from the condition
$\partial_\Lambda H =0$.
Assuming these fluctuations are small, one may  linearize
Eq.~(\ref{eq:eq_of_motion_Lambda}) near its equilibrium point.
Being transformed to the frequency representation,
such linearized matrix equation of motion takes the form
\begin{eqnarray}
  \label{eq:eq_of_motion_Y_linear}
 \left(-\frac{i\omega}{2} +\mathsf{\Gamma}^{-1}\right)  \Lambda (\omega)
&=& \xi(\omega)\,,
\end{eqnarray}
where  $\mathsf{\Gamma}$ is the Hessian matrix of the second derivatives of
the impurity Hamiltonian \emph{at fixed  momentum $P$}. Its matrix elements are
\begin{eqnarray}
  \label{eq:sec_der}
  \mathsf{\Gamma}^{-1} =\begin{pmatrix}
\Gamma_{++} & \Gamma_{+-} \\
\Gamma_{-+} & \Gamma_{--}
\end{pmatrix}^{-1}=
\left.
\begin{pmatrix}
H_{\Lambda_+\Lambda_+} & H_{\Lambda_+\Lambda_-} \\
H_{\Lambda_-\Lambda_+} & H_{\Lambda_-\Lambda_-}
\end{pmatrix}\right|_{\Lambda= \Lambda^{(0)}}
\end{eqnarray}
Solution of  equation (\ref{eq:eq_of_motion_Y_linear}) takes the form
\begin{equation}
\label{eq:perturbative solution}
    \Lambda(\omega)=\left(1-\frac{i\omega }{2}\mathsf{\Gamma}\right)^{-1}\,\mathsf{\Gamma}
    \,\xi(\omega) =\mathsf{\Gamma}\,\xi(\omega) +
    \frac{i\omega}{2}\mathsf{\Gamma}^2\,\xi(\omega)+\ldots \,,
\end{equation}
where we consider it as a perturbative sequence in frequency, to avoid
spurious instabilities mentioned in Section~\ref{sec:coupling_phonons}.
Substituting this solution into the right hand side of
Eq.~(\ref{eq:momentum_diss-Lambda}) and averaging it over the Gaussian noise
(\ref{eq:noise-correlator}), one finds for the momentum loss rate
\begin{eqnarray}
  \label{eq:friction0}
   \dot P=F_\mathrm{fr}=
-\frac{1}{2}\int \frac{d\omega}{2\pi} \,\,\omega^2\,
\mbox{Tr} \Big\{\big\langle\xi(\omega)\xi^\dagger(\omega)\big\rangle
\big( \mathsf{\Gamma}^\dagger \mathsf{V}^{-1}\mathsf{\Gamma}
-\mathsf{V}^{-1}\mathsf{\Gamma}^\dagger \mathsf{\Gamma} \big)\Big\}=
-\frac{1}{2}\int \frac{d\omega}{2\pi} \,\,\omega^3\, \mbox{Tr}
\Big\{  \mathsf{F} (\omega)  \big[ \mathsf{\Gamma}^\dagger, \mathsf{V}^{-1}\big]
\mathsf{\Gamma}\Big\} \, ,
\end{eqnarray}
where we kept only leading order in frequency. The matrix-valued function
$\mathsf{F} (\omega)$ is odd in frequency, selecting only
odd powers of $\omega$  from the expression it is multiplied by.
Equation~(\ref{eq:friction0}) provides the expression for
the viscous friction force  acting on the
impurity from the normal component of the liquid. It can be identified with
the Raman two-phonon scattering mechanism
\cite{Gangardt09,Gangardt2010Quantum,Muryshev2002Dynamics,LandauKhalatnikov1949ViscosityI,LandauKhalatnikov1949ViscosityII}.

Substituting the explicit form of the Ohmic noise correlator
({\ref{eq:matrices}) and performing the frequency integration,
one finds for the friction force
\begin{eqnarray}
  \label{eq:friction2}
  F_\mathrm{fr} =
-\big|{\Gamma}_{+-}\big|^2\frac{c\, }{c^2-V^2}
\int\!\frac{ d\omega}{2\pi} \,\,\omega^3
\left(\coth\frac{\omega}{2T_-}-\coth\frac{\omega}{2T_+}\right) =
-\frac{16\pi^3}{15}\big| {\Gamma}_{+-}\big|^2 \frac{T^4 }{c^2}
  \left(\frac{c^2+V^2}{c^2-V^2}\right) V\,.
\end{eqnarray}
The $T^4$ dependence of the friction force
was reported in \cite{castro96} and is
a direct consequence of the phase space for two-phonon scattering.
It can be also viewed as a one-dimensional  version of the
Khalatnikov-Landau result
\cite{LandauKhalatnikov1949ViscosityI,LandauKhalatnikov1949ViscosityII}
for the viscosity of liquid helium. It should be noted that the friction
force (\ref{eq:friction2}) is proportional to the off-diagonal matrix element 
$\Gamma_{+-}$ which represents backscattering of phonons by the depleton. It
depends on the momentum $P$ and, consequently velocity $V$ of the mobile 
impurity in addition to the parameters of the impurity-background
interactions. Below we derive the general expression for the 
backscattering amplitude and relate it to the equilibrium
dispersion $E(P,n)$ of the depleton.

\subsection{Backscattering amplitude }
\label{sec:backscattering}

The detailed information about microscopic impurity-liquid
interactions is contained in
the off-diagonal matrix element $\Gamma_{+-}$ which represents the effective
vertex for two-phonon scattering.  In the spirit of our phenomenological
approach it may be expressed,
as explained in  Appendix \ref{sec:matrix_gamma},
through partial derivatives of the equilibrium values
$\Phi_0(V,n)$ and  $N_0(V,n)$,
\begin{eqnarray}
  \label{gamma12_Vn}
  {\Gamma}_{+-} (V,n)&=& -\frac{\kappa m}{ M^*} \left[
    \frac{M-mN}{m n}\left(\frac{\der \Phi_0}{\der V}  +
      m\frac{\der  N_0}{\der n} \right)
    +\left(\frac{\der\Phi_0}{\der V}\frac{\der N_0}{\der n} -
    \frac{\der\Phi_0}{\der n}\frac{\der N_0}{\der V }\right)\right].
\end{eqnarray}

Treating  collective variables $\Phi_0=\Phi_0(P,n)$ and
$N_0=N_0(P,n)$ as function of
momentum rather than velocity and calculating the corresponding
derivatives as explained in Appendix~\ref{sec:matrix_gamma}
we find an alternative representation
\begin{eqnarray}
\label{gamma12_Pn}
{\Gamma}_{+-} (P,n) &=& -\frac{1}{c} \left(
\frac{M}{m}\frac{\der \Phi_0}{\der P}+  \Phi_0\frac{ \der N_0}{\der P}
- N_0\frac{\der \Phi_0}{\der P}
 +\frac{\der N_0}{\der n}\right).
\end{eqnarray}
The functions $\Phi_0$ and $N_0$ are obtained from the equilibrium dispersion
$E(P,n)$ as explained in Section \ref{sec:dyn_eq} (see, \emph{e.g.}
Eq.~(\ref{eq:Phi N})). This relations can be used to obtain an
equivalent representation for the backscattering amplitude ${\Gamma}_{+-}$ 
given by Eq.~(\ref{eq:Gamma12_derE}) in Appendix~\ref{sec:matrix_gamma}.  

\begin{figure}[t]
\centering
\includegraphics[width=0.6\columnwidth]{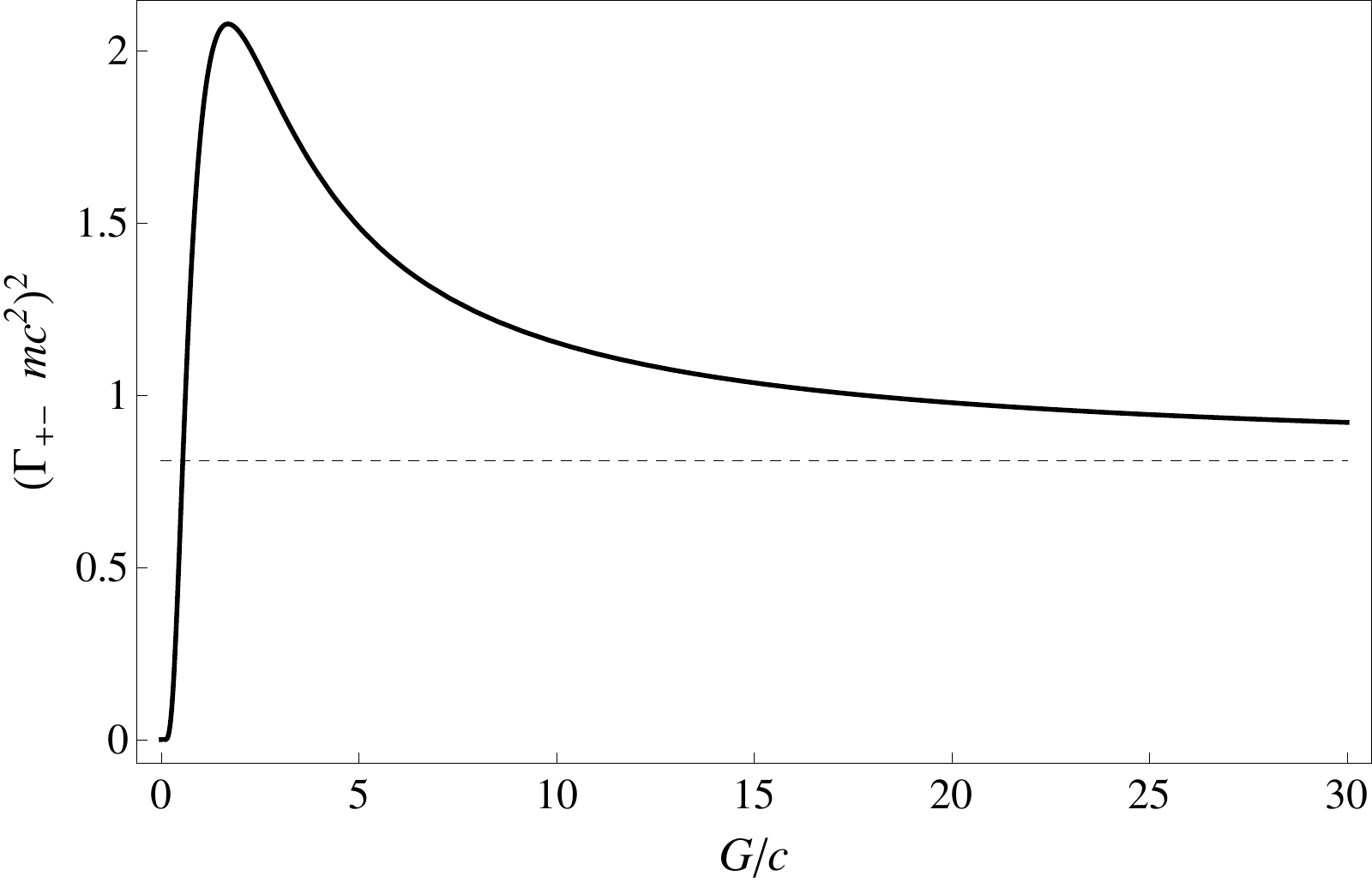}
\caption{$V=0$ and $\Phi=0$
(\emph{i.e.}, $P=0$) dissipation rate as a function of
 the impurity coupling $G$ for $\kappa=10$ and $M=m$ (solid line).
The dashed line corresponds
to the strong coupling $G\gg c$ limit, Eq.~(\ref{eq:gamma_strong_coupling}).}
\label{fig:gamma12_big_G}
\end{figure}

Two remarks are in order. First, 
it should be noticed that the backscattering amplitude
given by  Eqs.~(\ref{gamma12_Vn}),~(\ref{gamma12_Pn}) 
depends  on the velocity $V$ or momentum $P$ of the 
depleton. While the momentum dependence is unambiguous, there can be
different regimes corresponding to the \emph{same} velocity characterized by 
different values of the superfluid phase~$\Phi$.
Second, it can be shown by calculating dispersion $E(P,n)$
by Bethe Ansatz  method that the off-diagonal matrix element
$\Gamma_{+-}$ vanishes identically for integrable  mobile impurities,
such as lowest excitation branch in Lieb-Liniger \cite{Lieb_Liniger_1963} or
Yang-Gaudin \cite{CN_Yang_1967,Gaudin_1967} models and the soliton of the Gross-Pitaevskii equation (see Appendix~\ref{sec:dark_soliton}). We postpone the
details of these exact  calculations to another publication.  
Below we use the semiclassical approach valid for the weakly
interacting bosonic background which provides results for 
backscattering amplitude
in a wide range of parameters  relevant to experiments.

\subsubsection{Strong coupling regime $G/c\gg 1$}

In the strong coupling regime the leading terms in the
the number of depleted particles, the second equation in 
(\ref{eq:V N for large G})  is  a constant,
$N_0=2\kappa\gg 1$ and  the momentum is dominated  
by the supercurrent and $P=n\Phi$.
These observation simplify greatly the expression (\ref{gamma12_Pn}) for
backscattering amplitude and one gets 
\begin{eqnarray}
  \label{eq:gamma_strong_coupling}
  \Gamma_{+-} =
  \frac{1}{mc^2}\left[1-\frac{M}{\kappa m}\right]\, ,
\end{eqnarray}
where we have used the fact that $\der \kappa/\der n = \kappa/2n$. 
The expression (\ref{eq:gamma_strong_coupling}) 
is momentum or velocity independent constant, which equals  
$1/mc^2$ for a not too heavy impurity $M\ll \kappa m$.

\subsubsection{Weak coupling regime $G/c\ll 1$ for slow impurity}

In the weak coupling regime the backscattering amplitude varies strongly with
the velocity of the depleton. Here we present results for slow impurity which 
can be found by evaluating $\Gamma_{+-}$ for $V=0$.

There are two distinct physical regimes
corresponding to a slow impurity: one, corresponding to
the solution with  $\Phi_0\simeq 0$ and momentum $P\simeq MV\simeq 0$
dominated by the bare impurity and one corresponding to $\Phi_0\simeq\pi$ and
momentum $P\simeq \pi n$ dominated by the background supercurrent.
For $\Phi\approx 0$ we have
\begin{equation}
\label{eq:V=0 Phi=0}
N_0(V,n)=2\kappa\left(1+\frac{G}{2c}-\sqrt{1+\left(\frac{G}{2c}\right)^2}\right)
\, ,\qquad
\Phi_0(V,n)=2\frac{V}{c}
\left(-1+\frac{G}{2c}+\sqrt{1+\left(\frac{G}{2c}\right)^2}\right),
\end{equation}
while near $\Phi=\pi$, we find
\begin{equation}
\label{eq:V=0 Phi=pi}
N_0(V,n)=2\kappa\, , \qquad \Phi_0(V,n)=\pi-2\frac{V}{c}\left(1+\frac{G}{c}\right)\, .
\end{equation}
These expressions  are correct up to terms
$\mathcal{O}(V^{2})$ for the number of particles and $\mathcal{O}(V^{3})$ for
the phase.

In the case  $\Phi_0\simeq 0$ we  use
Eqs.~(\ref{eq:V=0 Phi=0}) to calculate $\Gamma_{+-}$.
For weak coupling  to the
leading order in both $g/c=1/\kappa$ and $G/c$  we have,
\begin{equation}
\label{eq:Gamma12 V=0 small G}
\Gamma_{+-}=\frac{1}{mc^{2}}\left(\frac{G}{c}\right)
\left(\frac{mG}{Mg}-1\right)\, ,
\qquad
G/c\ll 1 .
\end{equation}
This result agrees with Eq.(5) in Ref.~\cite{Gangardt09} and indicates
that friction vanishes near integrable point  $M=m$ and $G=g$,
at least for small velocities near $\Phi=0$. 
The results  for arbitrary coupling $G/c$ are presented  
in Figs.~\ref{fig:gamma12_big_G}.

In the case $\Phi_0\simeq\pi$, we use Eqs.(\ref{eq:V=0 Phi=pi}) and  calculate
effective mass $M^*=M-mN+n \der_V \Phi=M-2\kappa m
-2\kappa m(1+G/c)$.  Using Eq.(\ref{gamma12_Vn}) we find for arbitrary coupling
\begin{equation}
\label{eq:Gamma12 V=0 Phi=pi}
\Gamma_{+-}=\frac{1}{mc^{2}}\frac{(M/\kappa m)
\left(1+2G/c\right) -2G/c}{M/\kappa m-2G/c-4},
\end{equation}
It comes as no surprise that Eq.(\ref{eq:Gamma12
V=0 Phi=pi}) vanishes in the limit $M,G\rightarrow0$, where one recovers the
dark soliton result from Appendix \ref{sec:dark_soliton}. Indeed, there it is mentioned that such soliton excitation  is transparent for
phonons due to its integrability and, consequently,
$\Gamma_{+-}=0$ for all velocities. In the limit of weak coupling Eq.(\ref{eq:Gamma12 V=0 Phi=pi})
becomes
\begin{eqnarray}
\label{eq:gamma12 limits1}
\Gamma_{+-}=\frac{1}{mc^2}\left(\frac{2G}{c}-\frac{M}{\kappa m}\right)\, .
\end{eqnarray}
At the Yang-Gaudin integrable point (\cite{CN_Yang_1967,Gaudin_1967},
see also \cite{zvonarev_2007}) of the \emph{quantum} system defined as
$M=m$ and $G=g$, the corresponding limit of the
second equation in (\ref{eq:gamma12
  limits1}) is proportional $g/c=1/\kappa$ which is a small parameter in the
semiclassical limit and therefore our calculation 
goes beyond the accuracy of the  approximations adopted above.

\subsection{Bloch oscillations in the presence of viscous friction}
\label{sec:BO_friction}

Imagine the impurity is dragged by the weak external
force $F$. 
There is  a velocity $V<V_\mathrm{c}$ such that $F +
F_\mathrm{fr} (V)=0$ and the system reaches a steady state with constant
drift velocity $V_\mathrm{D}$ and \emph{no} oscillations, 
$V_\mathrm{B}=0$. 
If the applied force is very small, the drift velocity can be found from the 
small velocity limit of Eq.~(\ref{eq:friction2}) in the linear response form,
$V_\mathrm{D}=\sigma_T  F$, where the finite temperature mobility is given by
\begin{eqnarray}
  \label{eq:mob_T}
  \sigma^{-1}_T = 
  \frac{16\pi^3}{15 }\frac{ T^4} {c^2} \big|\Gamma_{+-}(V=0) \big|^{2}.
\end{eqnarray}
in accordance with \cite{castro96,Gangardt09}. Increasing the external force,
leads to a slightly nonlinear dependence of the drift velocity on $F$ (due to
the non-linear velocity dependence of $F_{\textrm{fr}}(V)$,
cf. Eq.~(\ref{eq:friction2})), until the the maximal possible velocity
$V_\mathrm{c}$ is reached for $F=F_\mathrm{min}= -F_\mathrm{fr}(V_c)$. Beyond
this critical force no steady state solution can be found and the impurity
performs Bloch oscillations along with the drift. In such a nonlinear regime
the amplitude, period of the oscillations, drift velocity, as well as the
momentum-dependent viscous backscattering amplitude, Eq.~(\ref{gamma12_Pn}),
are controlled by the equilibrium dispersion relation.

We illustrate the above scenario using the model of a strongly coupled
impurity.  In this case the critical velocity is small and backscattering
amplitude is velocity-independent. The critical force is then found from the
linear response as $F_\mathrm{min} =V_\mathrm{c}/\sigma_T$.  We can use the
relation $V=V_\mathrm{c}\sin\Phi+\kappa \dot{\Phi}/2n$ (i.e., the second of
Eqs.~(\ref{eq:eq. of motion 2})) and the fact that $P=n\Phi$ to write down the
equation of motion for the superfluid phase,
\begin{eqnarray}
  \label{eq:phase_strong}
  \dot \Phi = \frac{F_\mathrm{min}}{n(1+\kappa/2\sigma_T n^2)}\Big(\lambda-\sin\Phi\Big)\, ,
\end{eqnarray}
where $\lambda =F/F_\mathrm{min}$. Introducing dimensionless time variable 
$s=(F_\mathrm{min}/n(1+\kappa/2\sigma_T n^2)) t$ and assuming, without loss of generality,
$\Phi(0)=0$, the solution of Eq.~(\ref{eq:phase_strong}) is found to be
\begin{eqnarray}
  \label{eq:phase_strong_solution}
  \tan\frac{\Phi}{2} =
  \frac{\lambda \tanh\left(\frac{\sqrt{1-\lambda^2}}{2}s\right)}
  {\sqrt{1-\lambda^2}+\tanh\left(\frac{\sqrt{1-\lambda^2}}{2}s\right)}
\end{eqnarray}
For $\lambda <1$ it describes velocity approaching its limiting value
$V_\infty=V_\mathrm{c}\sin\Phi (\infty) = \lambda V_\mathrm{c}$.  
For $\lambda > 1$ Eq.~(\ref{eq:phase_strong_solution}) describes a periodic function with period
\begin{eqnarray}
  \label{eq:period_strong}
  \tau_\mathrm{B} = \frac{t}{s}\frac{2\pi}{\sqrt{\lambda^2-1}}=
      \frac{2\pi n(1+\kappa/2\sigma_T n^2)}{\sqrt{F^2-F^2_\mathrm{min}}}\, .
\end{eqnarray}
The corresponding drift velocity for $F>F_{\textrm{min}}$ can be found by
averaging the momentum relation $n\dot{\Phi}=F-V/\sigma_T$ over a single
period with the result

\begin{equation}
\label{eq:velocity_strong}
V_\mathrm{D}=V_c\left(\frac{F}{F_{\mathrm{min}}}-\frac{\sqrt{F^2/F_{\mathrm{min}}^2-1}}{1+\kappa/2\sigma_T n^2}\right).
\end{equation}

For $F_{\mathrm{min}}\ll F\ll F_{\mathrm{max}}$, one may expand
Eq.~(\ref{eq:velocity_strong}) to obtain a small temperature correction
\begin{equation}
\label{eq:velocity_strong2}
V_{\mathrm{D}}/c\approx\frac{F}{F_{\mathrm{max}}}+\frac{c}{V_c}\frac{F_{\mathrm{min}}}{F}\left(\frac{1}{2}V_c^2/c^2-(F/F_{\mathrm{max}})^2\right),
\end{equation}
which gives the previous result Eq.~(\ref{eq:mob_large_G}) in the limit of
small $T$, or more specifically, small $F_{\mathrm{min}}$. Interestingly,
Eq.~(\ref{eq:velocity_strong2}) predicts an \emph{increase} in the drift with
increasing $T$ if $F/F_{\mathrm{max}}\lesssim V_c/c$. This suggests that for a
small enough force, a significant \emph{drop} in the drift velocity may occur
as the system is cooled below the critical temperature, when the impurity
enters the regime of Bloch oscillations, see Fig.~\ref{fig:VD_vs_T}.

\begin{figure}[t]
\centering
\includegraphics[width=0.6\columnwidth]{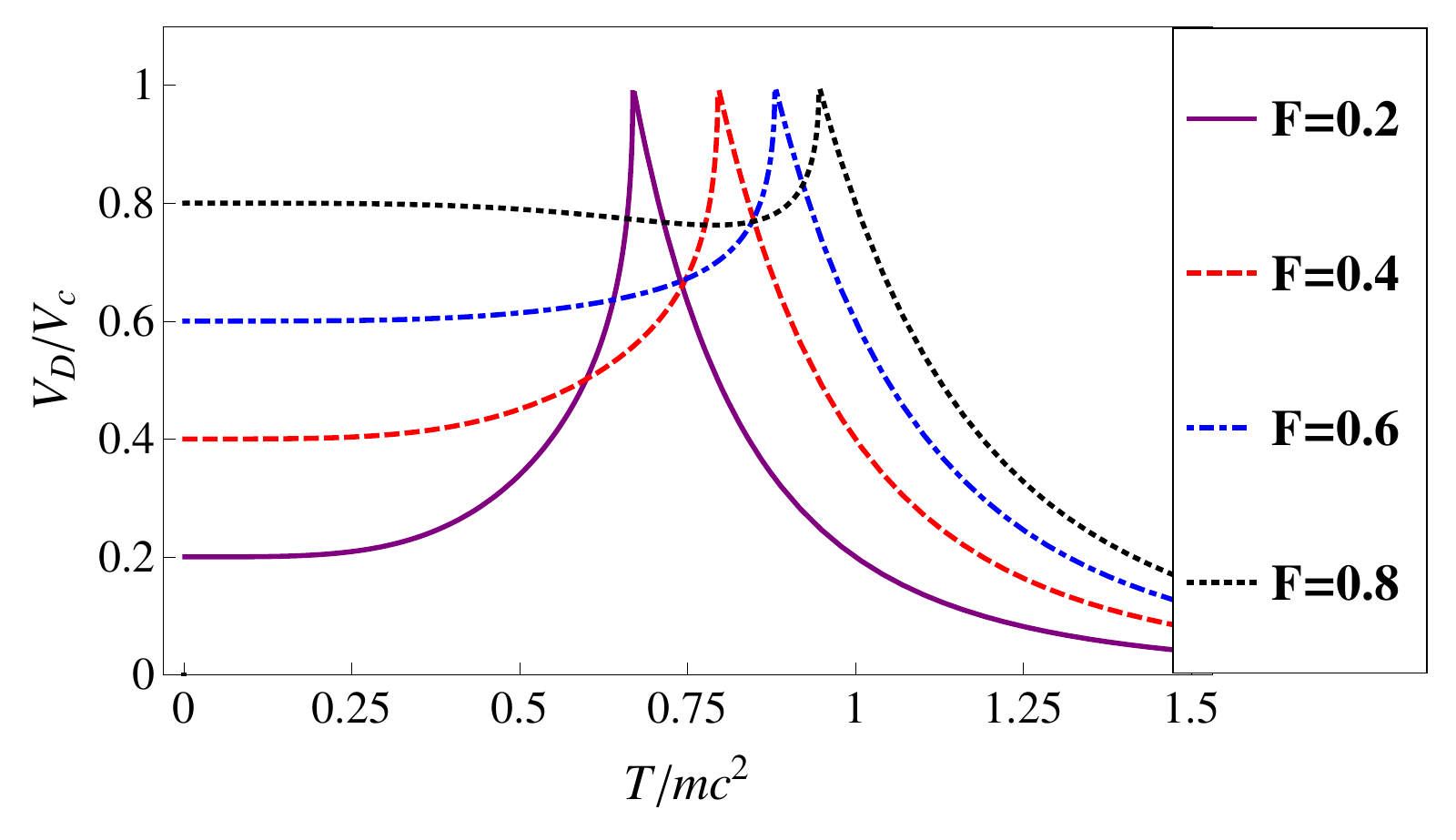}
\caption{Drift velocity of a strongly coupled impurity in a weakly interacting
  bose liquid as a function of temperature $T$ for various forces as predicted
  by Eq.~(\ref{eq:velocity_strong}). $F$ is given in units of
  $F_{\mathrm{max}}V_c/c$, while the Luttinger parameter is taken to be
  $K=\pi\kappa=8\pi^4/15\approx52$ so that, to a good approximation,
  $F_{\mathrm{min}}(T=mc^2)=F_{\mathrm{max}}V_c/c$.}
\label{fig:VD_vs_T}
\end{figure}

The above analysis demonstrates, \emph{inter alia}, a non-monotonic dependence
of the drift velocity on the parameter $F/F_{\mathrm{min}}$ as it is increases
past 1 and enters the regime of Bloch oscillations. This may occur either by
fixing the temperature and increasing the force, or by fixing the external
force and cooling the system.  In contrast to the vanishing amplitude
$V_\mathrm{B}$ near $F_\mathrm{max}$, indicated by the second equation in
(\ref{eq:drift_and_amp}), it is rather the divergent period,
Eq.~(\ref{eq:period_strong}) for $F$ approaching $F_\mathrm{min}$ from above
which leads to the disappearance of Bloch oscillations.

\section{Discussion of the results\label{sec:discussion}}

\begin{figure}[t]
\centering
\includegraphics[width=0.6\columnwidth]{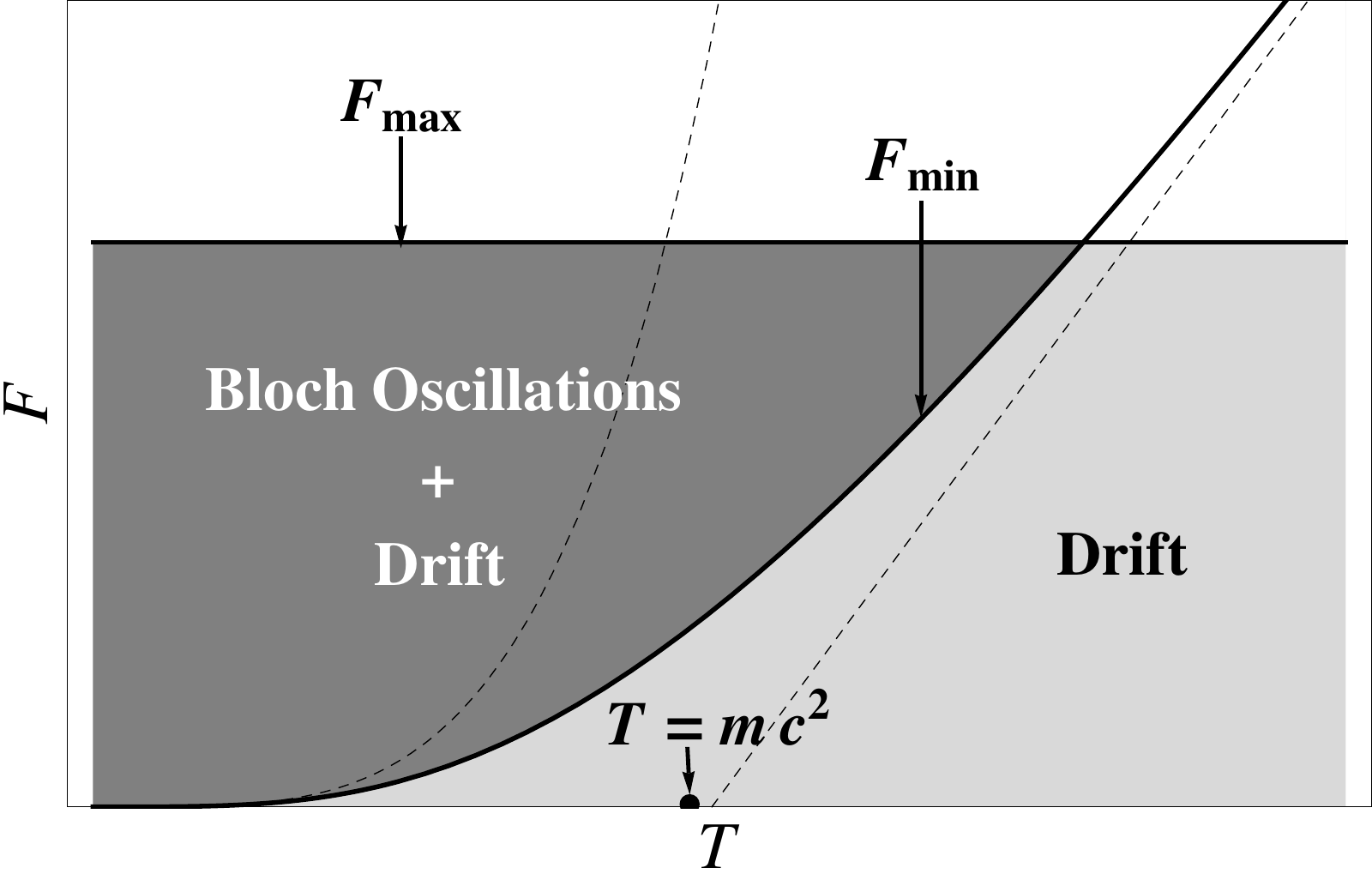}
\caption{Schematic force vs. temperature diagram of impurity motion in a 1D
  quantum liquid. In the region $F<F_{\textrm{min}}$ (light gray) the impurity
  drifts with mobility $\sigma_T\propto T^{-4}$ for $T\ll mc^2$. For $T\gg
  mc^2$, the mobility scales $\sigma_T\propto T^{-1}$ (dashed are the small
  and large temperature asymptotics). For
  $F_{\textrm{min}}<F<F_{\textrm{max}}$ (dark gray) one has Bloch oscillations
  + drift. Above $F_{\textrm{max}}$, our theory is inapplicable and we expect
  some kind of incoherent acceleration, possibly corresponding to a supersonic
  impurity which has \emph{escaped} its self-induced depletion cloud.}
\label{fig:F_T phase plane}
\end{figure}

The coupling with phonons governed by the universal term,
Eq.~(\ref{eq:imp-phonons}), results in transfer of energy and momentum between
the depleton and the background. At zero temperature it takes the form of
radiation of phonons by accelerated impurity similar to the radiative damping
in classical electrodynamics \cite{Ginzburg_Applications_of_Electrodynamics}
and leads to the finite mobility of the depleton.  For sufficiently weak
forces the mobility, defined as the ratio of the drift velocity to the applied
force, see Eq.~(\ref{eq:V drift2}) can be expressed via equilibrium
dispersion, Eq.~(\ref{eq:mob}) reminiscent of the Kubo linear response
theory. The drift, however, is superimposed with the essentially
\emph{non-linear} Bloch oscillations of the velocity. The amplitude of the
oscillations is shown to vanish as the external force attains the upper
critical force $F_\mathrm{max}$ reflecting the limit of validity for the
description based on the scale separation between the healing length and the
phonon wavelength.

At finite temperature the thermal phonons present in the system are scattered
by the depleton leading to the viscous friction force,
Eq.~(\ref{eq:friction2}). This in turn leads to the appearance of the lower
critical force $F_\mathrm{min}(T)\sim T^4$ which sets the lower limit for the
external force driving the Bloch oscillation.  In contrast to the situation in
the vicinity of $F_\mathrm{max}$, the approach to $F_\mathrm{min}$ makes the
Bloch oscillation disappear through the divergence of the period,
Eq.~(\ref{eq:period_strong}).  Below $F_\mathrm{min}$ the system enters the
regime of non-oscillatory drift characterized by the velocity $V_\mathrm{D} <
V_\mathrm{c}$, where all the momentum provided by the external force is
dissipated into phononic bath.

The viscous force contains valuable information about the fine details of the
interactions between particles. In this work we have confirmed and extended
the earlier observation \cite{Gangardt09}, that the backscattering amplitude,
given in terms of equilibrium properties by
Eqs.~(\ref{gamma12_Vn}),~(\ref{gamma12_Pn}) vanishes if the depleton is an
elementary excitation of an integrable model. This includes the dark soliton
excitation of Lieb-Liniger model \cite{Lieb_Liniger_1963} as well as spinons
in bosonic Yang-Gaudin model \cite{CN_Yang_1967,Gaudin_1967}. The microscopic
mechanism responsible for the absence of dissipation is due to destructive
quantum interference between various two-phonon processes. It can be traced
back to the absence of three-body processes lying at heart of integrability in
one dimension.  In contrast, the radiative processes due to the presence of
external force are always present in the dynamics of the depleton. This is
because the external potential is not, in general, compatible with the
integrability.

Various dynamical regimes of depleton are summarized in diagram
Fig.~\ref{fig:F_T phase plane}.  At sufficiently low temperatures there is a
wide parametric window $F_\mathrm{min}<F<F_\mathrm{max}$ for the external
force in which the Bloch oscillations can be observed experimentally. At
temperatures higher than chemical potential the mobility of the particle
becomes inversely proportional to temperature
\cite{Muryshev2002Dynamics,Gangardt2010Quantum} and moderates the growth of
$F_\mathrm{min}$ (see Fig.~\ref{fig:F_T phase plane}). The above range of
forces can be increased further by exploring the dependence of
$F_\mathrm{min}$ and $F_\mathrm{max}$ on the interaction parameters. In
particular, at integrable point $F_\mathrm{min}$ vanishes for \emph{any}
temperature.

Our approach to the dynamics of depleton is essentially classical, the quantum
mechanics enters only via parameters of the effective action.  The equations
of motion, Eqs.~(\ref{eq:eq_of_motion_Lambda}) neglect therefore the quantum
and thermal fluctuations of the collective variables of the depleton.  These
fluctuations can be taken into account by either simulating the Langevin
equation in the presence of equilibrium noises or by writing appropriate
Fokker-Planck equation for the distribution function. One of the most
important consequence of the fluctuations is expected to be the smearing the
boundaries between dynamical regimes in Fig.~\ref{fig:F_T phase plane}. We
leave this as well as the question about the role of quantum fluctuation for
further investigation.  Another important extension of the present work would
be the studies of the finite size effects due to the trap geometry relevant
for experiments with ultracold atoms.

\section{Acknowledgments}

We are indebted to L. Glazman, A. Lamacraft, M. Zvonarev, I. Lerner and
J.M.F.~Gunn for illuminating discussions. 
MS and AK were supported by DOE contract
DE-FG02-08ER46482. DMG acknowledges support by EPSRC Advanced Fellowship
EP/D072514/1. DMG and AK are thankful to Abdus Salam ICTP in Trieste for
hospitality at the early stages of this work.

\appendix
\section{Derivation of dissipative action}
\label{sec:s_diss}

Gaussian integration of the interaction term Eq.~(\ref{eq:s_int_keldysh_v})
using the following phononic propagators
\begin{eqnarray}
  \label{eq:dra}\nonumber
-i\Big\langle \chi^{\phantom{\dagger}}_\mathrm{cl} (x,t)
\chi^\dagger_\mathrm{cl} (x',t')\Big\rangle
&=& \mathsf{D}^K (x-x',t-t'),\qquad
-i\Big\langle \chi^{\phantom{\dagger}}_\mathrm{q} (x,t)
\chi^\dagger_\mathrm{cl} (x',t')\Big\rangle
= \mathsf{D}^A (x-x',t-t')
 \\
-i\Big\langle \chi^{\phantom{\dagger}}_\mathrm{cl} (x,t)
\chi^\dagger_\mathrm{q} (x',t')\Big\rangle
&=& \mathsf{D}^R (x-x',t-t'),\qquad
-i\Big\langle \chi^{\phantom{\dagger}}_\mathrm{q} (x,t)
\chi^\dagger_\mathrm{q} (x',t')\Big\rangle
= 0
\end{eqnarray}
leads to the quadratic nonlocal action
\begin{eqnarray}
  \label{eq:seff-lambda-delta}
S_\mathrm{eff}
&=& \frac{1}{2}\int\! \mathrm{d}t\mathrm{d}t^{\prime}\; \dot
\Lambda^\dagger_\mathrm{cl} (t) \; 
\der_t\Big[\mathsf{\Delta}^R(t-t')-\mathsf{\Delta}^A(t-t')\Big]
\left[\Lambda_\mathrm{q}(t^\prime) +\mathsf{V}^{-1}
    \dot \Lambda_\mathrm{cl}(t^\prime) X_\mathrm{q}(t^\prime)\right] \nonumber \\
    &-& \frac{1}{2}\int\! \mathrm{d}t \mathrm{d}t'
    \,\left[\Lambda^\dagger_\mathrm{q}(t)
      + X^{\phantom{\dagger}}_\mathrm{q}(t)\dot\Lambda^\dagger_\mathrm{cl}(t)\mathsf{V}^{-1}\right]
    \der^2_t\mathsf{\Delta}^K(t-t')
    \left[\Lambda_\mathrm{q}(t') +
      \mathsf{V}^{-1}\dot\Lambda_\mathrm{cl}(t')X_\mathrm{q}(t')\right]\, ,
\end{eqnarray}
where $\mathsf{\Delta}^{R,A,K} (t) = \mathsf{D}^{R,A,K}(Vt,t)$ are phonon
propagators restricted to the impurity trajectory.

Inverting the matrix in Eq.~(\ref{eq:dinverse}) and taking appropriate
analytic structure in the complex $\omega$ plane we obtain Fourier components
of the retarded and advanced propagators,
\begin{eqnarray}
  \mathsf{D}^{R}(q,\omega)=\Big[ \mathsf{D}^{A}(q,\omega)\Big]^\dagger
=\int\!dxdt\;
e^{-iqx+i\omega t} \,\mathsf{D}^{R} (x,t) =
 \begin{pmatrix}
     \frac{1}{q(\omega-cq+i0)} & 0\\
    0 & -\frac{1}{q(\omega+cq+i0)}
    \end{pmatrix}\, .
\end{eqnarray}
This leads to
\begin{eqnarray}
  \label{eq:dr-da}
  \mathsf{D}^{R}(q,\omega) - \mathsf{D}^{A}(q,\omega) = \frac{2\pi}{iq}
\begin{pmatrix}
     \delta(\omega-cq) & 0\\
    0 & -\delta(\omega+cq)
    \end{pmatrix}
\end{eqnarray}
and, subsequently,
\begin{eqnarray}
  \label{eq:delta_t}
  \der_t \Big[\mathsf{\Delta}^R(t)-\mathsf{\Delta}^A(t)\Big] = \frac{1}{2\pi}\int \mathrm{d} q\,
  \mathrm{d} \omega\, e^{i(qV-\omega) t}
  \left(\frac{qV-\omega}{q}\right)
  \begin{pmatrix}
     \delta(\omega-cq) & 0\\
    0 & -\delta(\omega+cq)
    \end{pmatrix} =- \begin{pmatrix}
     \delta(t) & 0\\
    0 & \delta(t)
    \end{pmatrix}
\end{eqnarray}


The noise terms, \emph{i.e.} the second line of
Eq.~(\ref{eq:seff-lambda-delta}) are controlled by the Keldysh component $D^K$
and its derivatives. Assuming thermal equilibrium of phonons in the laboratory
frame and using Fluctuation-Dissipation Theorem we have
\begin{eqnarray}
  \label{eq:phonon_propagator_K}
  \mathsf{D}^{K}(q,\omega) &=& \coth\left(\frac{\omega}{2T}\right)
  \Big(\mathsf{D}^{R}(q,\omega)-\mathsf{D}^{A}(q,\omega)\Big) \, .
\end{eqnarray}
Using Eq.~(\ref{eq:dr-da}) we find the Fourier component of the Keldysh
propagator restricted to the classical trajectory
\begin{eqnarray}
  \label{eq:delta_RAK}\nonumber
  \Delta^K (\omega) &=&
  \int \frac{\mathrm{d} q}{2\pi}\,\coth\left(\frac{\omega+qV}{2T}\right)
  \Big(\mathsf{D}^{R}(q,\omega+qV)-\mathsf{D}^{A}(q,\omega+qV)\Big)=
  \frac{1}{i\omega}\begin{pmatrix}
     \coth\frac{\omega}{2T}\frac{c}{c-V} & 0\\
     0 &  \coth\frac{\omega}{2T}\frac{c}{c+V}\end{pmatrix},
\end{eqnarray}
Substituting  its Fourier transform
it into Eq.~(\ref{eq:seff-lambda-delta}) and taking
double time derivative leads to the second term in Eq.~(\ref{eq:seff-lambda}).

\section{Grey Solitons}
\label{sec:dark_soliton}
Dynamic properties of one-dimensional bosons of mass $m$
weakly interacting via repulsive short range
potential proportional to coupling constant $g$
can be studied within the Gross-Pitaevskii
description using  the following Lagrangian
\begin{eqnarray}
  \label{eq:GP_lagrangian}
  L = \int \mathrm{d}x\left( i\bar{\psi}\der_t\psi -
  \frac{1}{2m} \left|\der_x\psi\right|^2 -
  \frac{g}{2} \left|\psi\right|^4 + \mu\left|\psi\right|^2\right)\, .
\end{eqnarray}
Here $\psi$ is the quasi-condensate wavefunction corresponding to the
asymptotic density $|\psi(\pm\infty,t)|^{2}=n$ and vanishing supercurrent at
infinity. The condition of weak interactions is $mg/n\ll1$.  Minimizing the
action defined by the Lagrangian Eq.~(\ref{eq:GP_lagrangian}) leads to the
Gross-Pitaevskii equation,
\begin{equation}
i\partial_{t}\psi=\left(-\frac{\partial_{x}^{2}}{2m}+
g|\psi|^{2}-\mu\right)\psi.
\label{eq:gross_pitaevskii_0}
\end{equation}
Substituting a constant solution $\psi_0 = \sqrt{n}$ one obtains the chemical
potential $\mu=gn=mc^2$ related to the sound velocity $c=\sqrt{gn/m}$. In
addition to uniform solution, Gross-Pitaevskii equation (\ref{eq:gross
  pitaevskii}) admits a one-parameter family of solutions
\begin{equation}
\psi_\mathrm{s}(x-Vt;\Phi_\mathrm{s})=
\sqrt{n}\left(\textrm{cos}\left(\frac{\Phi_\mathrm{s}}{2}\right)-
i\textrm{sin}\left(\frac{\Phi_\mathrm{s}}{2}\right)\textrm{tanh}
\left(\frac{x-Vt}{l}\right)\right),\label{eq:wavefunction_no_impurity}
\end{equation}
known as grey solitons \cite{PitaevskiiStringariBook,Tsuzuki_1971}.  They can
be visualized as a dip moving with velocity $V$ and having a core size
$l=1/m\sqrt{(c^2-V^2)}$.  The solution,
Eq.~(\ref{eq:wavefunction_no_impurity}) is characterized by the total phase
drop and number of expelled particles related to the velocity as
\begin{eqnarray}
\label{eq:ds_characteristics}
  \Phi_\mathrm{s} (V,n) &=&  2 \arccos\frac{V}{c},\qquad
N_\mathrm{s}(V,n) = \int\mathrm{d} x\, \left(n-|\psi_\mathrm{s}(x)|^2\right)
= \frac{2}{g}\sqrt{c^2-V^2}  \, .
\end{eqnarray}
Momentum and energy of the soliton are given by
\cite{Tsuzuki_1971,Shevchenko1988,Konotop2004Landau}
\begin{eqnarray}
  \label{eq:momentum_energy_soliton}
 P_\mathrm{s}=n\Phi_\mathrm{s}-mN_\mathrm{s}V\, ,\qquad
E_\mathrm{s}=\frac{4}{3} cn
\sin^3(\Phi_\mathrm{s}/2) = \frac{mg^2}{6}N_\mathrm{s}^3
\end{eqnarray}
which allows to define
the Lagrangian
\begin{eqnarray}
  \label{eq:lagr_difference}
  L_\mathrm{s} (V,n) =
 (n\Phi_\mathrm{s}-mN_\mathrm{s}V) V -
  \frac{mg^2}{6}N_\mathrm{s} ^3,
\end{eqnarray}
Using soliton Lagrangian (\ref{eq:lagr_difference}) and putting $M=0$
in Eqs.~(\ref{eq:dLdV}) and (\ref{eq:dLdn}) one sees
immediately that the variables in (\ref{eq:ds_characteristics})
coincide with the collective variables $\Phi$, $N$.
Therefore the grey soliton can be viewed as a model for
a \emph{massless} impurity consisting of depletion cloud only.

We invert Eqs.~(\ref{eq:ds_characteristics}) which yields
\begin{eqnarray}
  \label{eq:ds_rhou}
  V_0 = \frac{gN}{2\tan\frac{\Phi}{2}},
  \qquad  n_0=\frac{mgN^2}{4\sin^2\frac{\Phi}{2}}\, .
\end{eqnarray}
Using Eq.~(\ref{eq:lagrangian_impurity}) together with
Eqs.~(\ref{eq:ds_characteristics}) and (\ref{eq:lagr_difference})
yields the internal energy, Eq.~(\ref{eq:h_dark_soliton}) of the soliton,
\begin{eqnarray}
 \label{eq:ds_energy}
  H_\mathrm{d} (N,\Phi) &=&
\frac{mg^2}{6}N^3+ N\left(\frac{mV_0^2}{2}-gn_0\right) =
\frac{mg^2N^3}{8}\left(\frac{1}{3} - \frac{1}{\sin^2\Phi/2}\right).
\end{eqnarray}
The matrix of second derivatives \emph{at the equilibrium solution} reads
\begin{eqnarray}
  \label{eq:ds_matrix}
  \mathsf{H} =\begin{pmatrix} H_{\Phi\Phi} & H_{\Phi N} \\
                              H_{N \Phi}   & H_{NN}\end{pmatrix}=
 \begin{pmatrix} -cn
      \frac{1+\cos^2\Phi/2}{\sin\Phi/2} & 2mc^2\frac{\cos\Phi/2}{\sin\Phi/2}
      \\
      2mc^2\frac{\cos\Phi/2}{\sin\Phi/2} &  -\frac{m^2c^3}{n}\frac{1+\cos^2\Phi/2}{\sin\Phi/2} \end{pmatrix}
\end{eqnarray}
The backscattering amplitude $\Gamma_{+-}$, calculated with the help of
Eq.~(\ref{eq:gamma12_h}), vanishes identically due to the integrability \cite{Note1} of the Gross-Pitaevskii
equation (\ref{eq:gross_pitaevskii_0}).

As it was shown in \cite{Fedichev1999Dissipative,Muryshev2002Dynamics}
a weak cubic nonlinearity $-(\alpha/6)|\psi|^6$
in the Lagrangian (\ref{eq:GP_lagrangian}) breaks the
integrability of the model. Cubic terms  describe three-body interactions
which arise from virtual transitions to higher transverse states
of tightly confined one-dimensional liquid \cite{Mazets2008Breakdown}.
Here we extent the results in
Ref.\cite{Gangardt2010Quantum} and calculate the amplitude
of the corresponding dissipation processes for the whole range of soliton
velocities.

The corresponding  correction
to the Lagrangian, Eq.~(\ref{eq:lagr_difference}) of the soliton can be
calculated to the leading order by evaluating it with the unperturbed
solution,
\begin{eqnarray}
  \label{eq:delta_lagr_soliton}
 \delta L_\mathrm{d} = -\frac{\alpha}{6}
 \int \left(|\psi_\mathrm{d}(x)|^2-n\right)^3  =
 \frac{8}{45}\frac{\alpha n^3}{mc}\left(1-\frac{V^2}{c^2}\right)^{5/2} =
  \frac{\alpha m^2g^2}{180}  N^5 = -  \delta H_\mathrm{d} .
\end{eqnarray}
Here we have used the expression Eq.~(\ref{eq:ds_characteristics}) for the
number of expelled particles $N$ to calculate the correction to the energy
relying on the theorem of  small increments.
The corresponding change in the matrix (\ref{eq:ds_matrix})
of second derivatives
\begin{eqnarray}
  \label{eq:delta_h}
  \mathsf{\delta H} =- \frac{\alpha m^2 g^2 }{9} N^3\times
  \begin{pmatrix} 0 & 0 \\ 0 & 1\end{pmatrix}
\end{eqnarray}
can be taken perturbatively in the calculation of the off-diagonal matrix
element $\delta\Gamma_{+-}$. We have
\begin{eqnarray}
  \label{eq:pert_gamma}
  \mathsf{\delta\Gamma} =-\mathsf{\Gamma} \left( \mathsf{T^{-1} \delta H
      \left({T^{-1}}\right)^\dagger } \right)  \mathsf{\Gamma}
  \, .
\end{eqnarray}
Substituting Eqs.~(\ref{eq:delta_h}), (\ref{eq:phipm}),  yields
\begin{eqnarray}
  \label{eq:pert_gamma_pm}
  \delta\Gamma_{+-} = \frac{\alpha\kappa m^2 g^2}{18} N^3\,
  \Gamma_{++}\Gamma_{--}  =\frac{\alpha\kappa m^2 g^2}{18}
  \frac{N^3}{\det\mathsf{H}}
  \end{eqnarray}
Here we have used the fact that
the matrix $\mathsf{\Gamma}$ is diagonal in the leading order in
$\alpha$ and $\det (\mathsf{T^\dagger T}) =1$.
%
The determinant of the matrix (\ref{eq:ds_matrix}) is
$\det \mathsf{H} = \left(m c^2 \sin \Phi/2\right)^2$.
Substituting it into Eq.~(\ref{eq:pert_gamma_pm}) and using
(\ref{eq:ds_characteristics}) leads to
\begin{eqnarray}
  \label{eq:gamma_soliton}
  \Gamma_{+-}=
-\frac{4}{9} \frac{\alpha n^2}{m^2c^4}\left(1- \frac{V^2}{c^2}\right)^{1/2}
\end{eqnarray}
in agreement with the results of Ref.\cite{Gangardt2010Quantum}.

\section{Impurity in a weakly interacting liquid}
\label{sec:equilibrium impurity}

To model the impurity coupled to a weakly interacting superfluid at $T=0$,
the Gross-Pitaevskii equation, Eq.~(\ref{eq:gross_pitaevskii_0})
is modified  in the presence of a delta function
potential moving with constant velocity $V$
\begin{equation}
i\partial_{t}\psi=\left(-\frac{\partial_{x}^{2}}{2m}+
g|\psi|^{2}-\mu+G\delta(x-Vt)\right)\psi.
\label{eq:gross pitaevskii}
\end{equation}
\begin{figure}[t]
\centering
\includegraphics[width=0.6\columnwidth]{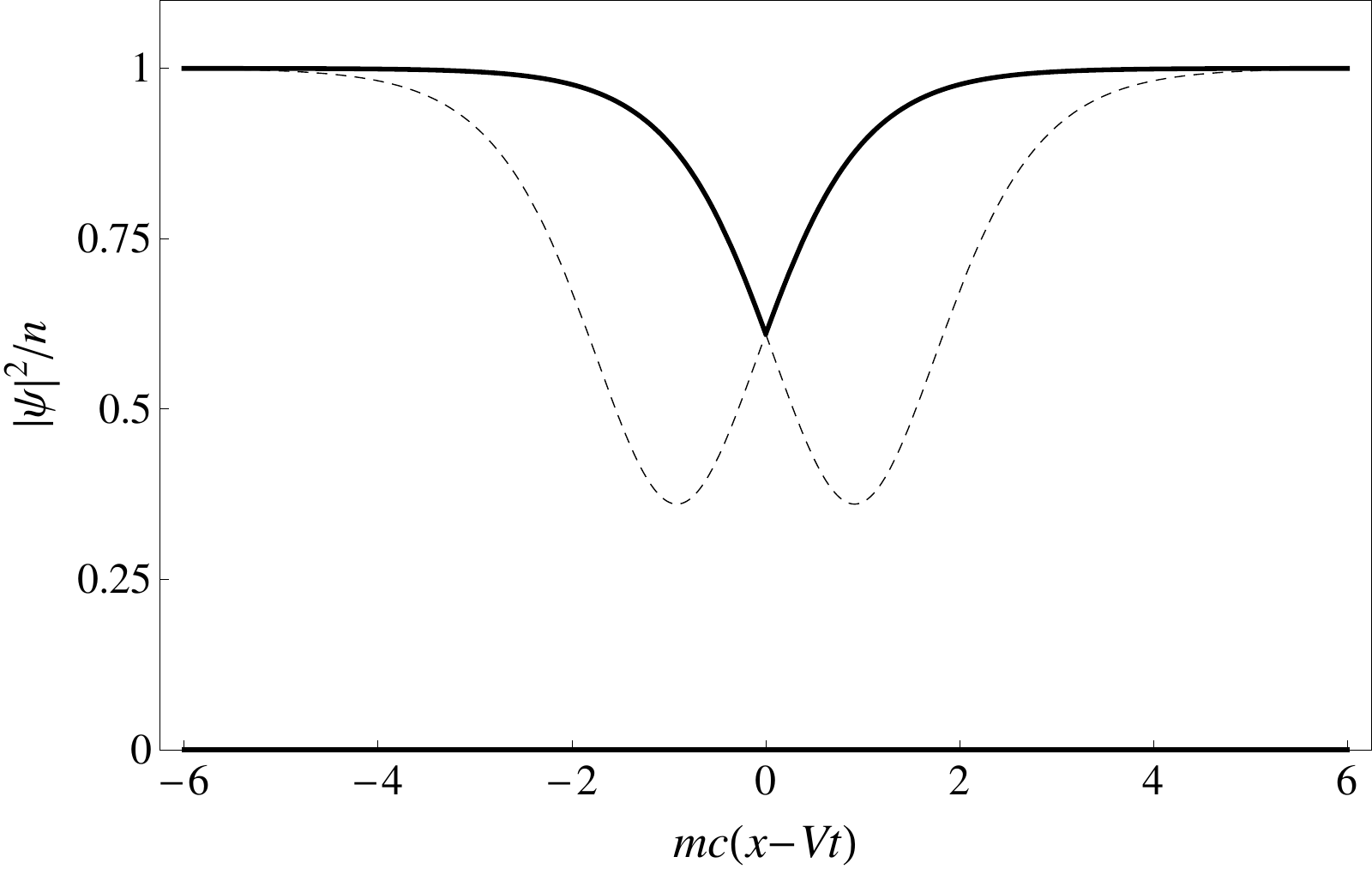}
\caption{Superfluid density in the presence of an impurity moving with
velocity $V/c=0.6$ and coupling $G/c=0.1$ (solid line). The dashed lines
corresponds to the double soliton solution Eq.(\ref{eq:wavefunction_no_impurity}) used to construct Eq.(\ref{eq:psi_with_impurity}).}
\label{fig:density}
\end{figure}
For $G>0$, the soliton solution, Eq.~(\ref{eq:wavefunction_no_impurity}) still
satisfies Eq.(\ref{eq:gross pitaevskii}) except at the location of the
impurity. Thus, one may construct a solution to Eq.(\ref{eq:gross pitaevskii})
by matching \emph{two} soliton solutions,
Eq.~(\ref{eq:wavefunction_no_impurity}) at the location of the impurity, as
shown in Fig.~\ref{fig:density}.  The proper solutions of Eq,~(\ref{eq:gross
  pitaevskii}) can thus be written
\begin{eqnarray}
\psi(y)=\begin{cases}
\psi_\mathrm{s}(y+x_{0};\Phi_\mathrm{s})e^{i\Phi_{0}/2}, & y>0\\
\psi_\mathrm{s}(y-x_{0};\Phi_\mathrm{s})e^{-i\Phi_{0}/2}, & y<0\end{cases}
\label{eq:psi_with_impurity}
\end{eqnarray}
where $y=x-Vt$ and velocity is always related as $V/c=
\cos(\Phi_\mathrm{s}/2)$ to the phase $\Phi_\mathrm{s}$ parameterizing the
soliton configuration, Eq.~(\ref{eq:wavefunction_no_impurity}).  The solutions
$\psi_\pm$ to the right (left) of the impurity satisfy the boundary
conditions:
$\psi_{+}(0)=\psi_{-}(0),\,\psi_{+}^{\prime}(0)-\psi_{-}^{\prime}(0)=2mG\psi(0)$.
Using (\ref{eq:psi_with_impurity}), the boundary conditions give rise to the
following two equations for $\Phi_{0}$ and $z=\textrm{tanh}(x_{0}/l)$
\begin{eqnarray}
\label{eq:z1}
\tan\left(\frac{\Phi_{0}}{2}\right)&=&z\,\tan
\left(\frac{\Phi_{s}}{2}\right)\\
\label{eq:z2}
\sin^{3}\left(\frac{\Phi_\mathrm{s}}{2}\right)\left(1-z^{2}\right)z
&=&\frac{G}{c}\left[\cos^{2}\left(\frac{\Phi_\mathrm{s}}{2}\right)
+z^{2}\sin^{2}\left(\frac{\Phi_\mathrm{s}}{2}\right)\right]\, .
\end{eqnarray}

Equations (\ref{eq:z1}),~(\ref{eq:z2}) permit a solution only for
$V<V_{c}(G/c)$ where $V_{c}$ is some critical velocity that depends only upon
the parameter $G/c$
\cite{Hakim_PhysRevE.55.2835,Gunn_Taras_PhysRevB.60.13139}. This can be seen
by considering the right and left hand sides of the second equation
(\ref{eq:z2}). While the left hand side is bounded by the
maximum at $z_\mathrm{max}=1/\sqrt{3}$ the right hand side grows quadratically
with $z$ and therefore solution exist only for a limited range of $\Phi_s$,
which leads to the above-mentioned limitation on velocity.

For this reason we choose
to parameterize the solution, Eq.(\ref{eq:psi_with_impurity}), by the
\emph{total} phase drop across the impurity, $\Phi=\Phi_{s}-\Phi_{0}$, which
happens to permit a solution for \emph{any} $\Phi$.
Thus, upon solving Eqs.~(\ref{eq:z1}),~(\ref{eq:z2}) one finds the relations
$z=z(\Phi,G/c)$ and $\Phi_\mathrm{s}=\Phi_\mathrm{s}(\Phi,G/c)$.  It can
easily be seen from Eqs.~(\ref{eq:z1}),~(\ref{eq:z2}) that these
functions are periodic in $\Phi$.
The number of expelled particles and momentum can be calculated and expressed
through the phase $\Phi$ as
\begin{eqnarray}
  \label{eq:nz}
  N=\int dx(n-|\psi|^{2})=2\kappa\,\sin\frac{\Phi_s}{2}(1-z);
  \qquad P=n\Phi+(M-mN)V,
\end{eqnarray}
The energy may also be
calculated and expressed in terms of $\Phi$ as
\begin{equation}
  \label{eq:energy_phi}
  E=\frac{1}{2}MV^{2}+
  \int dx\left[\frac{|\psi_{x}|^{2}}{2m}+\frac{g}{2}(n-|\psi|^{2})^{2}\right]
  +G|\psi(0)|^{2}=\frac{1}{2}MV^{2}+\frac{4}{3}nc\,
  \sin^{3}\frac{\Phi_{s}}{2}\left[1-\frac{3}{4}z-\frac{1}{4}z^{3}\right].
\end{equation}
Alternatively, we may solve for $\Phi=\Phi(P)$ by inverting the second of
Eqs.~(\ref{eq:nz}). Substituting it into the energy function,
Eq.~(\ref{eq:energy_phi}) one obtains the dispersion relation $E(P)$ plotted
in Fig.~\ref{fig:dispersion} for the impurity in a weakly interacting bose
liquid.

In the weak coupling regime $G/c\ll 1$  the critical velocity can be
obtained by neglecting the $z$-dependence of the r.h.s. Eq.~(\ref{eq:z2}).
This is justified \emph{a posteriori} as
at the solution  $\Phi_\mathrm{s} \ll 1$. Expanding the
trigonometric functions and using $z_\mathrm{max}=1/\sqrt{3}$
in the l.h.s  one obtains $(\Phi_\mathrm{s}/2)^3=(3\sqrt{3}/2)\,G/c$, which
justifies our approximation. We thus have the critical velocity
\begin{eqnarray}
  \label{eq:vc}
  \frac{V_\mathrm{c}}{c} = 1-\frac{1}{2}\left(\frac{\Phi_\mathrm{s}}{2}\right)^2=
  1-\frac{3}{4}\left(\frac{\sqrt{2}G}{c}\right)^{2/3}\, .
\end{eqnarray}
Solving Eqs.~(\ref{eq:z1}),~(\ref{eq:z2}) for arbitrary velocity or momentum
is cumbersome and we resort to numerical methods.

In the strong coupling  limit $G\gg c$ we
determine the dependence of $z$ and $\Phi_\mathrm{s}$ on $\Phi$ to order
$c/G$. This
is done by writing $\Phi_\mathrm{s}\simeq\pi+\frac{c}{G}\Phi_\mathrm{s}^{(1)}$
and $z\simeq\frac{c}{G}z^{(1)}$ and finding coefficients
$\Phi_\mathrm{s}^{(1)}$ and
$z^{(1)}$ from Eqs.~(\ref{eq:z1}),~(\ref{eq:z2}). We have
\begin{equation}
z(\Phi)\simeq\frac{c}{G}\textrm{cos}^{2}\frac{\Phi}{2};\qquad\Phi_{s}(\Phi)\simeq\pi-\frac{c}{G}\sin\Phi\,.
\label{eq:z large G}\end{equation}
Using Eqs.~(\ref{eq:z large G}),~(\ref{eq:nz}) one has
\begin{equation}
V\simeq\frac{c^{2}}{2G}\sin\Phi;\qquad
N\simeq 2\kappa\left(1-\frac{c}{G}\textrm{cos}^{2}\frac{\Phi}{2}\right).
\label{eq:V N for large G}
\end{equation}
The first equation has the Josephson form with
the critical velocity $V_\mathrm{c} =c^2/2G$ in the strong coupling limit.
It is also clear from the second equation that in the leading approximation
$N$ is a constant $N\simeq 2\kappa$. This results in  the Josephson form for the
energy, Eq.~(\ref{eq:josephson}).

At $V=0$ the equations (\ref{eq:z1}),~(\ref{eq:z2}) simplify considerably by
putting $\Phi_\mathrm{s}=\pi$. There are two solutions,
\begin{equation}
\label{eq:z V=0}
z_{0}=\sqrt{1+\left(\frac{G}{2c}\right)^{2}}-\frac{G}{2c}\, , \qquad
z_{\pi}=0\, ,
\end{equation}
The $z_0$ root
corresponds to $\Phi=0$ and describes a background only slightly perturbed by
the stationary impurity.  The $z_\pi$ root corresponds to the dark soliton
solution with $\Phi=\pi$, which persists for $G>0$ because the density
vanishes at the impurity location \emph{i.e.}, 
there is no additional energy cost to
put the impurity in the center of a dark soliton.

\section{
Solution of equation of motion for a strongly coupled impurity}
\label{sec:solution_strong}

We write the sinusoidal forms for the velocity and the phase of depleton
\begin{equation}
  V(t)=V_{\mathrm{D}}+V_{\mathrm{B}}\sin(\omega_{\mathrm{B}}t+\delta_{V});\qquad
  \dot{\Phi}(t)=\omega_{\mathrm{B}}(1+A\cos(\omega_{\mathrm{B}}t+\delta_{\Phi})),
  \label{eq:2}
\end{equation}
depending on \emph{a priori} unknown
parameters $V_{\mathrm{D}},V_{\mathrm{B}},\omega_{\mathrm{B}},\delta_{V},\delta_{\Phi},A$ and substitute
them into Eq.~(\ref{eq:eq. of motion 2}). One
first arrives at the following relations between the time \emph{independent}
components.
\begin{equation}
  n\omega_{\mathrm{B}}=F-\frac{1}{2}\frac{\kappa V_{\mathrm{D}}\omega_{\mathrm{B}}^{2}}{c^{2}-V_{\mathrm{D}}^{2}};\qquad
  \kappa\omega_{\mathrm{B}}/2n=V_{\mathrm{D}}\implies\omega_{\mathrm{B}}/2mc^{2}=V_{\mathrm{D}}/c
  \label{eq:3}
\end{equation}
Solving these equations one obtains $V_{\mathrm{D}}$ and $\omega_{\mathrm{B}}$ in
Eq.~(\ref{eq:drift_and_amp}).
In the limit $F\ll F_{\textrm{max}}=2nmc^2$
we recover the linear dependence $V_{\mathrm{D}}/c=F/F_{\textrm{max}}$. The
leading order deviation from linearity is given by
\begin{equation}
  V_{\mathrm{D}}/c\approx\frac{F}{F_{\mathrm{max}}}
  \left(1-\frac{F^{2}}{F_{\mathrm{max}}^{2}}\right).
  \label{eq:5}
\end{equation}
Substituting the ansatz (\ref{eq:2})
into the first of Eqs.~(\ref{eq:eq. of motion 2})
gives
a relation between the time \emph{dependent} components,
\begin{equation}
  A\left[n\omega_{\mathrm{B}}+\frac{\kappa\omega_{\mathrm{B}}^{2}V_{\mathrm{D}}}{c^{2}-V_{\mathrm{D}}^{2}}\right]
  \cos(\omega_{\mathrm{B}}t+\delta_{\Phi})+V_{\mathrm{B}}(M-mN)\omega_{\mathrm{B}}
  \left[\cos(\omega_{\mathrm{B}}t+\delta_{V})+\frac{1}{2}\frac{\kappa\omega_{\mathrm{B}}}{M-mN}
    \frac{c^{2}+V_{\mathrm{D}}^{2}}{(c^{2}-V_{\mathrm{D}}^{2})^{2}}
    \sin(\omega_{\mathrm{B}}t+\delta_{V})\right]=0,
  \label{eq:6}
\end{equation}
where we neglected terms $\mathcal{O}(A^{2})$, $\mathcal{O}(V_{\mathrm{B}}^{2})$
and $\mathcal{O}(AV_{\mathrm{B}})$ to keep the calculation to first order
in $V_{c}/c$, as both $V_{\mathrm{B}}$ and $A$ will be seen to scale with
$V_{c}/c$. The second term in brackets in Eq.~(\ref{eq:6}) is simplified
employing the formula
\begin{equation}
  \cos x+\alpha\sin x
  =-\sqrt{1+\alpha^{2}}\cos(x+\pi/2+
\arctan(1/\alpha)).
\label{eq:7}
\end{equation}
In order to cancel the term $\propto\textrm{cos}(\omega_{\mathrm{B}}t+\delta_{\Phi})$,
we require a definite relation between the phases and amplitudes.
The constraint for the phase is
$\delta_{V}+\pi/2+\textrm{arctan}(1/\alpha)=\delta_{\Phi}$.
From Eq.~(\ref{eq:6}) we have
\begin{equation}
  \alpha=\frac{1}{2}\frac{\kappa\omega_{\mathrm{B}}}{M-mN}\frac{c^{2}
    +V_{\mathrm{D}}^{2}}{(c^{2}-V_{\mathrm{D}}^{2})^{2}}
  =\frac{m\kappa}{M-mN}\frac{1+V_{\mathrm{D}}^{2}/c^{2}}{(1-V_{\mathrm{D}}^{2}/c^{2})^{2}}
  \frac{V_{\mathrm{D}}}{c}.
\label{eq:8}
\end{equation}
The equality of amplitude implies a relation between $A$ and $V_{\mathrm{B}}$,
namely
\begin{equation}
  A=\frac{V_{\mathrm{B}}}{c}\frac{M-mN}{m\kappa}
  \sqrt{1+\alpha^{2}}\left(\frac{1-V_{\mathrm{D}}^{2}/c^{2}}
    {1+V_{\mathrm{D}}^{2}/c^{2}}\right).
  \label{eq:9}
\end{equation}
Finally, we substitute the ansatz into the second of Eqs.~(\ref{eq:eq. of
  motion 2})
to obtain
\begin{equation}
  \left[AV_{\mathrm{D}}+\frac{\alpha V_{\mathrm{B}}}
    {\sqrt{1+\alpha^{2}}}\right]
  \cos(\omega_{\mathrm{B}}t+\delta_{\Phi})+
  \frac{V_{\mathrm{B}}}{\sqrt{1+\alpha^{2}}}
  \sin(\omega_{\mathrm{B}}t+\delta_{\Phi})
  =-V_{c}\sin(\omega_{\mathrm{B}}t+\Phi_{0}),
  \label{eq:10}
\end{equation}
where we used
$\textrm{sin}(\omega_{\mathrm{B}}t+\delta_{V})
=-\frac{1}{\sqrt{1+\alpha^{2}}}\left(\textrm{sin}(\omega_{\mathrm{B}}t+\delta_{\Phi})
+\alpha\textrm{cos}(\omega_{\mathrm{B}}t+\delta_{\Phi})\right)$
and $\Phi_{0}$ is some initial phase of $\Phi$ which comes from integrating
the second of Eqs.~(\ref{eq:2}). Using Eq.~(\ref{eq:7}) again finally gives
\begin{equation}
  V_{\mathrm{B}}=V_{c}\frac{1-V_{\mathrm{D}}^{2}/c^{2}}
  {\sqrt{1+V_{\mathrm{D}}^{4}/\alpha^2c^{4}}}
  =V_{c}\frac{1-V_{\mathrm{D}}^{4}/c^{4}}
  {\sqrt{\left(1+V_{\mathrm{D}}^{2}/c^{2}\right)^{2}
      +\left(\frac{M-mN}{\kappa m}\right)^{2}
      \left(1-V_{\mathrm{D}}^{2}/c^{2}\right)^{4}V_{\mathrm{D}}^{2}/c^{2}}},
  \label{eq:VB_theory}
\end{equation}
where $V_{\mathrm{D}}$ is given by Eq.~(\ref{eq:drift_and_amp}). As expected, $V_{\mathrm{B}}$ is an even
function of $F$ since $V_{\mathrm{D}}$ is odd. For small $F\ll F_{\textrm{max}}$
Eq.~(\ref{eq:VB_theory}) gives Eq.~(\ref{eq:drift_and_amp}).

\section{Calculation of the backscattering amplitude
$\Gamma_{+-}$ }
\label{sec:matrix_gamma}

Using the transformation Eq.~(\ref{eq:yupsilon}) we have
$\mathsf{\Gamma} =
\mathsf{T^\dagger} \mathsf{H}^{-1}\mathsf{ T}$ which leads to  
\begin{eqnarray}
  \label{eq:gamma12_h}
{\Gamma}_{+-} = =\frac{1}{\det\mathsf{H}}
 \left(\kappa H_{NN}- \frac{1}{\kappa} H_{\Phi\Phi}\right)\, .
\end{eqnarray}
The matrix of second
derivatives
\begin{eqnarray}
  \label{eq:h}
\mathsf{H}=
\begin{pmatrix}
H_{\Phi\Phi} & H_{\Phi N} \\
H_{N\Phi} & H_{NN}
\end{pmatrix}
  \end{eqnarray}
is calculated at equilibrium values of  $\Phi$ and  $N$. 
Differentiating  Eq.~(\ref{eq:hamiltonian_nophonons})
and taking into account Eq.~(\ref{eq:velocity_phi_n})
for the dependence of the velocity $V$ on $\Phi$ and $N$ at constant
momentum $P$, one can rewrite Eq.~(\ref{eq:h}) as
\begin{eqnarray}
  \label{eq:sec_der_1}
  \mathsf{H} =
\frac{1}{M-mN} \begin{pmatrix} n^2 & -nmV \\ -nmV & m^2 V^2 \end{pmatrix}
-\frac{1}{\det\mathsf{\Omega}}\begin{pmatrix}
    N_\mu & -\Phi_\mu   \\
    -N_j & \Phi_j
  \end{pmatrix}
\end{eqnarray}
The last term in this equation represents the matrix of second derivatives
of $H_\mathrm{d} (\Phi,N)$. Owing to properties of Legendre transformation
we have expressed it as  the inverse of the  Hessian matrix,
\begin{eqnarray}
  \label{eq:omega}
 \mathsf{\Omega}=\begin{pmatrix}
    \Omega_{jj} & \Omega_{j\mu }  \\
    \Omega_{\mu j} & \Omega_{\mu\mu}
  \end{pmatrix} =\begin{pmatrix}
    \Phi_j & \Phi_\mu   \\
    N_j & N_\mu
  \end{pmatrix}
\end{eqnarray}
of the thermodynamical potential $\Omega_\mathrm{d}'(j',\mu')$.
Hereafter we  drop the primes over $\Omega$,  $j$ and $\mu$ for clarity.
In writing Eq.~(\ref{eq:omega}) we
used Eq.~(\ref{eq:thermo_der}) to express double derivatives
as  derivatives of equilibrium values $\Phi$ and $N$ with
respect to underlying values of  the supercurrent and chemical potential.

Using relations  $mV=\der\mu/\der V$, $n=-\der j/\der V$ in the first term of
Eq.~(\ref{eq:sec_der_1}) simplifies considerably
the determinant
\begin{eqnarray}
  \label{eq:deth}
  \det\mathsf{H} = \frac{1}{\det\mathsf{\Omega}}
\left(1-\frac{1}{M-mN}\left[m^2V^2N_\mu + n^2\Phi_j -
mnV(N_j+\Phi_\mu)\right]\right) =
\frac{1}{\det\mathsf{\Omega}} \frac{M^*}{M-mN}.
\end{eqnarray}
by using the  effective mass Eq.~(\ref{eq:eff_mass})
of the equilibrated impurity.
We use this fact and invert the matrix in Eq.~(\ref{eq:sec_der_1}) by a
standard  procedure and find for the off-diagonal matrix element
 \begin{eqnarray}
\label{gamma12_jm}
\Gamma_{+-} = \frac{1}{ \kappa M^*} \Big[(M-mN)
  \left(\kappa^2\Phi_j- N_\mu \right) +
  n^2\left(1-V^2/c^2\right)\left(\Phi_j N_\mu - \Phi_\mu N_j\right)\Big].
\end{eqnarray}

In the heavy particle limit $M\simeq M^*\gg m$ only the first term in the
square bracket survives. In this limit
$\Gamma_{+-}$ can be
identified with backscattering amplitude of a static impurity
$\Gamma_{+-}= \Gamma_{\infty}= \kappa \Phi_j-\kappa^{-1} N_\mu=
-\left(\der_V\Phi+m\der_n N\right)/mc$. To obtain this result  
we inverted the
relations in Eq.~(\ref{eq:linear_derivatives}),
\begin{eqnarray}
  \label{eq:linear_derivatives_inv}
  \begin{pmatrix} \der_j \\ \der_\mu \end{pmatrix} &=&
  \frac{1}{m\left(c^2-V^2\right)}
  \begin{pmatrix} -mc^2/n  & mV \\ -V & n  \end{pmatrix}
   \begin{pmatrix} \der_V \\ \der_n\end{pmatrix},
\end{eqnarray}
and  used the identity $n\der_n\Phi - mV\der_n N =
V\der_V\Phi-(mc^2/n)\der_V N$. The latter is 
 obtained from equality of the
mixed derivatives of the Lagrangian $L(V,n)$, obtained by
differentiating Eqs.~(\ref{eq:dLdV}),~(\ref{eq:dLdn}) by $n$ and
$V$ respectively.
The second term in the square brackets in Eq.~(\ref{gamma12_jm}) is
transformed by applying (\ref{eq:linear_derivatives_inv}). Combining it with
the  term, $\Gamma_\infty \,(M-mN)/M^*$ yields Eq.~(\ref{gamma12_Vn}).

To deal with equilibrium functions  $\Phi(P,n)$,
$N(P,n)$ obtained for a given momentum $P$ rather than
velocity $V$ we use the fact that
$\der_V = M^*\der_P$ and the
following  relation for the derivatives with respect to the density
\begin{eqnarray}
  \label{eq:dern}
  \big(\der/\der n \big)_V &=& \big(\der/\der n \big)_P -
  M^* \left(\der V/\der n\right)_P \der_P\, .
\end{eqnarray}
Differentiating Eqs.~(\ref{eq:dLdV}),(\ref{eq:dLdn}) with respect to $P$ and
$n$ at constant momentum leads to
\begin{eqnarray}
  \label{eq:der2EP}
  \begin{pmatrix} \der_P \Phi \\ \der_P N \end{pmatrix} &=&
  \frac{1}{m\left(c^2-V^2\right)}
  \begin{pmatrix} mc^2/n  & mV \\ V & n  \end{pmatrix}
   \begin{pmatrix} (M^*-M+mN)/M^* \\ \der_n V+\Phi/M^*\end{pmatrix}\\
  \label{eq:der2En}
  \begin{pmatrix} \der_n \Phi \\ \der_n N \end{pmatrix} &=&
  \frac{1}{m\left(c^2-V^2\right)}
  \begin{pmatrix} mc^2/n  & mV \\ V & n  \end{pmatrix}
   \begin{pmatrix} -\Phi-(M-mN)\der_n V \\ \Phi\der_n V+(\alpha_d-\alpha
     N)\end{pmatrix},
\end{eqnarray}
where $\alpha_d = \der^2_n E$.
Substituting Eq.~(\ref{eq:dern})  into Eq~(\ref{gamma12_Vn}),
using the effective mass, Eq.~(\ref{eq:eff_mass}) together with
Eq.~(\ref{eq:der2EP})   we arrive at
Eq.~(\ref{gamma12_Pn}).  Alternatively one may wish to express $\Gamma_{+-}$
directly in terms of derivatives of the dispersion law $E(P,n)$ with the
result,
\begin{eqnarray}
  \label{eq:Gamma12_derE}
  \Gamma_{+-} = -\frac{1}{cn}\frac{1}{1-V^2/c^2} \left[\frac{M-mN}{m}
    +\frac{n^2}{mc^2} (\alpha_d -\alpha N)
 -\frac{(M-mN)^2-m^2\kappa^2\Phi^2}{mM^*} +2 m\kappa^2\Phi\der_n V\right]
\end{eqnarray}
In the limit of small velocities terms proportional to $\Phi$ can be neglected
and this expression becomes  is identical to Eq.~(3) in
Ref.\cite{Gangardt09} with $N=1-\delta_l$.


\end{document}